\documentclass[apj]{emulateapj}
\usepackage{multirow}
\usepackage{graphics}
\usepackage{rotating}
\usepackage{enumitem}
\usepackage{amsmath}
\usepackage{txfonts}
\usepackage{hyperref}
\usepackage{setspace}
\usepackage{float}
\usepackage{tablefootnote}
\usepackage{footnote}
\usepackage{threeparttable}
\usepackage[mediumspace,mediumqspace,Grey,squaren]{SIunits}

\hypersetup{
			hyperfootnotes=true,
			colorlinks=true,
			linkcolor=blue,
			citecolor=blue,
			urlcolor=blue,
			breaklinks=false,
}

\shorttitle{Galaxy Structure and Quenching at 0.6 $\lesssim$ \lowercase{$z$} $\lesssim$ 1.2}
\shortauthors{Kim et al.}

\begin{document}

\title{Galaxy Structure, Stellar Populations, and Star Formation Quenching at 0.6 $\lesssim$ \lowercase{$z$} $\lesssim$ 1.2}
\author{Keunho Kim\altaffilmark{1}, Sangeeta Malhotra \altaffilmark{2}, James E. Rhoads\altaffilmark{2}, Bhavin Joshi\altaffilmark{1}, Ignacio Fererras\altaffilmark{3}, and Anna Pasquali\altaffilmark{4}}

\altaffiltext{1}{School of Earth $\&$ Space Exploration, Arizona State University, Tempe, AZ 85287; Keunho.Kim@asu.edu}
\altaffiltext{2}{NASA Goddard Space Flight Center, Greenbelt, MD 20770}
\altaffiltext{3}{Mullard Space Science Laboratory, University College London, Holmbury St Mary, Dorking, Surrey RH5 6NT, UK}
\altaffiltext{4}{Astronomisches Rechen-Institut, Zentrum fuer Astronomie der Universitaet Heidelberg, Moenchhofstrasse 12-14, 69120 Heidelberg, Germany}

\def\OI{[\mbox{O\,{\sc i}}]~$\lambda 6300$}
\def\OIII{[\mbox{O\,{\sc iii}}]~$\lambda 5007$}
\def\OIIIs{[\mbox{O\,{\sc iii}}]~$\lambda 4363$}
\def\OIIIab{[\mbox{O\,{\sc iii}}]$\lambda\lambda 4959,5007$}
\def\SIIab{[\mbox{S\,{\sc ii}}]~$\lambda\lambda 6717,6731$}
\def\SII{[\mbox{S\,{\sc ii}}]~$\lambda \lambda 6717,6731$}
\def\NII{[\mbox{N\,{\sc ii}}]~$\lambda 6584$}
\def\NIIb{[\mbox{N\,{\sc ii}}]~$\lambda 6584$}
\def\NIIa{[\mbox{N\,{\sc ii}}]~$\lambda 6548$}
\def\NI{[\mbox{N\,{\sc i}}]~$\lambda \lambda 5198,5200$}

\def\OIIa{[\mbox{O{\sc ii}}]~$\lambda 3726$}
\def\OIIb{[\mbox{O{\sc ii}}]~$\lambda 3729$}
\def\NeIIIa{[\mbox{Ne{\sc iii}}]~$\lambda 3869$}
\def\NeIIIb{[\mbox{Ne{\sc iii}}]~$\lambda 3967$}
\def\OIIIa{[\mbox{O{\sc iii}}]~$\lambda 4959$}
\def\OIIIb{[\mbox{O{\sc iii}}]~$\lambda 5007$}
\def\HeII{{He{\sc ii}}~$\lambda 4686$}
\def\ArIVa{[\mbox{Ar{\sc iv}}]~$\lambda 4711$}
\def\ArIVb{[\mbox{Ar{\sc iv}}]~$\lambda 4740$}
\def\NIa{[\mbox{N{\sc i}}]~$\lambda 5198$}
\def\NIb{[\mbox{N{\sc i}}]~$\lambda 5200$}
\def\HeI{{He{\sc i}}~$\lambda 5876$}
\def\OI{[\mbox{O{\sc i}}]~$\lambda 6300$}
\def\OIb{[\mbox{O{\sc i}}]~$\lambda 6364$}
\def\SIIa{[\mbox{S{\sc ii}}]~$\lambda 6716$}
\def\SIIb{[\mbox{S{\sc ii}}]~$\lambda 6731$}
\def\ArIII{[\mbox{Ar{\sc iii}}]~$\lambda 7136$}

\def\Ha{{H$\alpha\,$}}
\def\Hb{{H$\beta\,$}}

\def\NIIHa{[\mbox{N\,{\sc ii}}]/H$\alpha$}
\def\SIIHa{[\mbox{S\,{\sc ii}}]/H$\alpha$}
\def\OIHa{[\mbox{O\,{\sc i}}]/H$\alpha$}
\def\OIIIHb{[\mbox{O\,{\sc iii}}]/H$\beta$}

\def\Ebmv{E($B-V$)}
\def\LOIII{$L[\mbox{O\,{\sc iii}}]$}
\def\Ledd{${L/L_{\rm Edd}}$}
\def\LOIIIs4{$L[\mbox{O\,{\sc iii}}]$/$\sigma^4$}
\def\LOIIIMbh{$L[\mbox{O\,{\sc iii}}]$/$M_{\rm BH}$}
\def\Mbh{$M_{\rm BH}$}
\def\Msigma{$M_{\rm BH} - \sigma$}
\def\Ms{$M_{\rm *}$}
\def\Msun{$M_{\odot}$}
\def\Msunyr{$M_{\odot}yr^{-1}$}
\def\bt{$B/T\,$}
\def\btr{$B/T_{\rm r}\,$}

\def\ergs{$~\rm ergs^{-1}$}
\def\kms{${\rm km}~{\rm s}^{-1}$}
\newcommand{\cms}{\mbox{${\rm cm\;s^{-1}}$}}
\newcommand{\pccm}{\mbox{${\rm cm^{-3}}$}}
\newcommand{\sauron}{{\texttt {SAURON}}}
\newcommand{\oasis}{{\texttt {OASIS}}}
\newcommand{\HST}{{\it HST\/}}

\newcommand{\Vg}{$V_{\rm gas}$}
\newcommand{\Sg}{$\sigma_{\rm gas}$}
\newcommand{\eg}{e.g.,}
\newcommand{\ie}{i.e.,}

\newcommand{\gandalf}{{\texttt {gandalf}}}
\newcommand{\fracDeV}{{\texttt {FracDeV}}} 
\newcommand{\ppxf}{{\texttt {pPXF}}}

\newcommand{\sersic}{S\'{e}rsic}

\begin{abstract}
We use both photometric and spectroscopic data from the {\it Hubble Space Telescope} to explore the relationships among 4000 \AA\ break (D4000) strength, colors, stellar masses, and morphology, in a sample of 352 galaxies with log$(M_{*}/M_{\odot}) > 9.44$ at 0.6 $\lesssim z \lesssim$ 1.2. We have identified authentically  quiescent galaxies in the $UVJ$ diagram based on their D4000 strengths. This spectroscopic identification is in good agreement with their photometrically-derived specific star formation rates (sSFR). Morphologically, most (that is, 66 out of 68 galaxies, $\sim$ 97 \%) of these newly identified quiescent galaxies have a prominent bulge component. However, not all of the bulge-dominated galaxies are quenched. We found that bulge-dominated galaxies show positive correlations among the D4000 strength, stellar mass, and the S\'ersic index, while late-type disks do not show such strong positive correlations. Also, bulge-dominated galaxies are clearly separated into two main groups in the parameter space of sSFR vs. stellar mass and stellar surface density within the effective radius, $\Sigma_{\rm e}$, while late-type disks and irregulars only show high sSFR. This split is directly linked to the `blue cloud' and the `red sequence' populations, and correlates with the associated central compactness indicated by $\Sigma_{\rm e}$. While star-forming massive late-type disks and irregulars (with D4000 $<$ 1.5 and log$(M_{*}/M_{\odot}) \gtrsim 10.5$) span a stellar mass range comparable to bulge-dominated galaxies, most have systematically lower $\Sigma_{\rm e}$ $\lesssim$ $10^{9}M_{\odot}\rm{kpc^{-2}}$. This suggests that the presence of a bulge is a necessary but not sufficient requirement for quenching at intermediate redshifts.

\end{abstract}	


\section{Introduction}
\label{sec:introduction}

How galaxies have formed and evolved over the Hubble time remains a challenging question in extragalactic astronomy. In particular, why and how galaxies have stopped their star formation activities during the course of their evolutionary paths is not clearly understood yet. Observational aspects have shown that galaxies largely form a bimodality in diverse diagrams (e.g., the Color-Magnitude diagram (CMD), the color-color diagram, and global star formation rate (SFR) versus stellar mass diagram), being separated into an actively star-forming group and a star-formation quiescent group. The former group forms the `blue cloud', while the later group forms a tight `red sequence' in the CMD. The intermediate parameter space, between the blue cloud and the red sequence, forms the `green valley' \citep{stra01,scha14,pand17,brem18,gu18}.

Previous studies \citep[e.g.,][]{pasq10,pengy10} have suggested that two main mechanisms quenched the star formation activity in galaxies. The first is mass-quenching, which indicates that the more massive galaxies are, the earlier they likely had formed stars and shut down star formation activities (i.e., ``downsizing'' scenario) \citep{cowi96,delu06,hain17}. The second is environmental quenching, which states that denser environments tend to make galaxies more passive, resulting in a larger fraction of quenched galaxies than in less dense regions \citep{pasq10,pengy10,khim15}.

Along with these two main mechanisms of quenching galaxies over a wide span of redshift, morphological properties of quiescent galaxies and star-forming counterparts show that quiescent galaxies tend to have a prominent bulge component, while most of star-forming galaxies tend to have prominent disk or clumpy structures \citep{oh13,huer16}. These morphologically-entangled features of quiescent and star-forming populations of galaxies have led to an idea of `morphological quenching' of galaxies \citep{mart09,deke09,mart13,deke14,genz14}. In addition to morphology, analyses of internal structure of galaxies based on galaxy structural parameters, such as the central surface density, a bulge-to-total ratio, and the S\'ersic index, also suggest that the presence of a prominent bulge component (or, similarly, compact central density in galaxy center) in galaxies, as the results of `inside-out' growth, is related to the quenching of galaxies. \citep{fran08,ilbe10,vand10,bell12,sain12,bluc14,lang14,tacc15,tacc16,lill16,huer16,jung17,whit17,willi17}.

Although the community seems to agree on the presence of a prominent bulge component as a useful indicator associated with  quenched galaxies, the mechanisms that are responsible for a bulge growth and the associated galaxy structure evolution have not been fully understood yet \citep[for a review]{cons14}. Recent studies \citep[e.g.,][]{marg16,huer15,huer16} have suggested that fractions of different morphological types of galaxies have evolved over cosmic time and the associated morphology-related dominant quenching mechanisms have changed accordingly. However, most of studies at intermediate or high redshifts \citep[e.g.,][]{huer16,owns16,hain17} have mainly focused on either optical photometric properties of galaxies with their morphology or spectroscopic and photometric properties of galaxies with a lack of morphological information.

In this study, we investigate the stellar properties of galaxies in the redshift range 0.6 $\lesssim$ \lowercase{$z$} $\lesssim$ 1.2 throughout the Hubble sequence \citep{hubb26,hubb36}, combining optical spectroscopic and photometric properties of galaxies as well as their visual morphological classification. The combination of most currently available optical diagnostics on galaxies enables us to comprehensively explore both the stellar and morphological properties of galaxies within the redshift range of interest, which is our hope in this study. Morphological behaviors in several diagrams such as a $U-V$ vs. $V-J$ color-color (hereafter, $UVJ$) diagram, the 4000 $\AA$ break \citep[i.e.,][]{bruz83,hami85} vs. stellar mass, and specific star formation rate (sSFR) vs. stellar mass are investigated.

This paper is organized as follows: in Section \ref{sec:sample}, we describe the observational data sets employed in this study.  In Section \ref{sec:data analysis}, we describe the 4000 $\AA$ break measurements, sample selection, and comparison of visually-classified morphology with other morphology indicators.  We present our results in Section \ref{sec:results}.  In Section \ref{sec:discussion},  we discuss the stellar and morphological properties of galaxies and the associated quenching mechanisms.  We close with a summary in Section~\ref{sec:summary}.

We adopt the $\Lambda$CDM cosmology of ($H_{0}$, $\Omega_{m}$, $\Omega_{\Lambda}$) = (70 $\rm{kms^{-1}}$ $\rm{Mpc^{-1}}$, 0.3, 0.7) wherever necessary. All magnitude in this paper are quoted in the AB system \citep{oke83}.

\section{The Observation data sets}
\label{sec:sample}

\subsection{The PEARS Survey}
\label{subsec:pears}
The 4000 $\AA$ break (hereafter, D4000) information for our sample galaxies is obtained from the Probing Evolution And Reionization Spectroscopically survey \citep[PEARS;][PI: S. Malhotra]{malh05}. The PEARS wide survey observed 8 fields, four each in the GOODS-North and GOODS-South \citep{giav04} regions. 
The slitless grism ACS/G800L instrument used for the survey has an average dispersion of 40 $\AA$ pixel$^{-1}$ and the spectral resolution R = 100 at 8000 $\AA$. The grism data are suitable for measuring the D4000 strength for galaxies within the redshift range of 0.6 $\lesssim z \lesssim$ 1.2, taking into consideration the observed wavelength coverage of the ACS/G800L instrument.

\subsection{Morphology Classification}
\label{subsec:morphology}
Galaxy morphology information is obtained from the CANDELS \citep{grog11,koek11} visual morphology classification catalog provided by \cite{kart15}. In that paper, the authors performed visual classifications on the four $HST$ bands (F606W, F850LP, F125W, and F160W) by voting for the most likely morphological types (among disk, spheroid, irregular/peculiar, compact/unresolved, and unclassifiable) for individual galaxies. Specifically, in the visual morphology classification procedure, the F160W band was primarily used and the other three bands were also used to cover different rest-frame wavelengths sensitive to different types of galaxy structures. Thus, the rest-frame wavelengths employed for the morphology classification range from near-UV ($\sim$ 3000 $\AA$) to near-IR ($\sim$ 8000 $\AA$) for most galaxies within the redshift range 0.6 $\lesssim$ \lowercase{$z$} $\lesssim$ 1.2. The four bands were simultaneously displayed when a classifier voted for morphologies for each galaxy. A minimum size of galaxy images is 84 pixels in each x and y-axis. Regarding voting for the likely morphologies of a galaxy, multiple choices of morphological types for a galaxy are allowed. Thus, the resulting morphology information from the catalog is given as the fractions of individual human classifiers who selected each of the several morphological types. In the following, we are concerned with the fractions of spheroid, disks, and irregulars, denoted as $f_{\rm sph}$, $f_{\rm disk}$, and $f_{\rm irr}$, respectively. Further details on the morphology classification procedure can be found in \cite{kart15}. 

Four types of morphology are considered in this work and classified as follows: spheroids satisfying the fractions of $f_{\rm sph} > 2/3$ and $f_{\rm disk} < 2/3$ and $f_{\rm irr} < 0.1$, early-type disks satisfying the fractions of $f_{\rm sph} > 2/3$ and $f_{\rm disk} > 2/3$ and $f_{\rm irr} < 0.1$, late-type disks satisfying the fractions of $f_{\rm sph} < 2/3$ and $f_{\rm disk} > 2/3$ and $f_{\rm irr} < 0.1$, and irregulars satisfying the fractions of $f_{\rm sph} < 2/3$ and $f_{\rm irr} > 0.1$. Additionally, regarding the morphology classification difference between early-type disks and late-type disks, we note that the only difference between the two morphological types is the value of $f_{\rm sph}$. That is, with the same morphological fractions of $f_{\rm disk} > 2/3$ and $f_{\rm irr} < 0.1$, if a galaxy has $f_{\rm sph} > 2/3 \ (< 2/3)$, the galaxy is classified as early (late)-type disks. Note our adopted morphological classification differs from the traditional approach that follows the Hubble tuning fork diagram, taking into account the spiral arm structure in late-type systems. Therefore, our definition of early-type disks is more likely to focus on the moderate dominance of both bulge and disk components, while our definition of late-type disks is more likely to focus on the dominance of a disk component compared to a relatively small (or no) bulge component, which does not consider the detailed spiral arms features as often adopted in the local Universe.

This classification scheme has been found effective in disks/spheroids separations, considering the close correlations between diskiness/bulginess, S\'ersic index $n$, and optical colors \citep{kart15,huer16}. Particularly, \cite{kart15} showed that their ``Mostly Disk'' galaxies have relatively lower S\'ersic indices with a mean of $\langle n\rangle = 1.01$ and bluer optical colors in the $UVJ$ diagram, while their ``Mostly Spheroid'' galaxies have relatively higher S\'ersic indices with a mean of $\langle n\rangle = 2.98$ and redder optical colors \citep[see their Figures 12 and 13]{kart15}. For our sample of galaxies, we will address the visual morphology and the comparison to other morphology indicators in detail in Section \ref{subsec:visual morphology vs CAS}.

The number of classifiers for our sample of galaxies ranges from 3 to 6. The mean number of classifiers is 4.09.

\subsection{Photometrically-derived Properties of Galaxies}
\label{subsec:stellar mass}
The photometrically-derived properties such as stellar mass, rest-frame colors, and star formation rate are obtained from the stellar mass catalog of \cite{sant15}, where different stellar masses derived by different spectral energy distribution (SED) fitting procedures are presented and compared. Their fitting is based on the 17 multi-wavelength photometric bands ranging from UV to mid-infrared for the GOODS-South field provided by \cite{guo13}.

Among the different stellar masses provided in the catalog, we adopt the stellar mass derived by the `Method 6a$_{\rm \tau}^{\rm NEB}$', in which Method 6a$_{\rm \tau}^{\rm NEB}$ is the team notation with the abbreviations of the SED fitting methods as designated in \cite{sant15}. Specifically, the Method 6a$_{\rm \tau}^{\rm NEB}$ makes use of the `zphot' code for the SED fitting based on the $\chi^{2}$ minimization fitting method \citep{gial98,font00}. Also, it adopts the Chabrier Initial Mass Function \citep{chab03} and the stellar templates from \cite{bruz03} with the exponentially-declining star formation history (so-called the direct-$\tau$ model, $\psi(t) \propto $ \rm{exp}$(-t/\tau)$). The resulting stellar mass and SFR estimates are provided with the least significant digit being 0.01 dex (base-10 log) and 0.01 decimal, respectively. The minimum SFR is set to 0.00 $M_{\odot} \rm {yr}^{-1}$.

Additionally, the Method 6a$_{\rm \tau}^{\rm NEB}$ includes the nebular emission component into its SED fitting procedures following the prescription of \cite{scha09}, where continuum emission is added to the stellar continuum, and emission lines are treated using the relative line intensities from \cite{ande03} for non-hydrogen lines such as He, C, N, O, and S, and from \cite{stor95} for hydrogen lines such as Ly$\alpha$ line and Balmer, Paschen, and Brackett series. Further details on the nebular emission prescription can be found in \cite{scha09}. However, more importantly, we note that the differences in stellar mass between those with and without the emission prescription are not significant. That is, the median value of the stellar mass differences is 0.00 $M_{\odot} \rm {yr}^{-1}$ for our sample of galaxies that will be described in Section \ref{subsec:sample selection} (see also Figure 3 and the discussion in Section 4.3 of \cite{sant15} on the nebular emission prescription for further details).

\subsection{Galaxy Structural Parameters}
\label{subsec:structural parameter}
Galaxy structural parameters for our sample are obtained from the publicly available galaxy structural parameters catalog provided by \cite{vand12}. The structural parameters in the catalog were derived by the two-dimensional light profile fitting procedures utilizing the GALFIT software for the CANDELS galaxies \citep{peng02,peng10}. A single S\'ersic component model \citep[i.e.,][]{sers68} was used to fit the two-dimensional light distribution of galaxies. For all of our sample galaxies, the structural parameters are measured with the $F160W$ image. Further details regarding the fitting procedures can be found in \cite{vand12}.

\section{Data analysis}
\label{sec:data analysis}

\subsection{D4000 Measurements}
\label{subsec:d4000 measurement}
The D4000 strength has been widely used as an optical regime age indicator for constituent stellar populations of galaxies \citep{bruz83,hami85,balo99,kauf03a,hath09,hain17}. It is defined by the average flux ratio of the red side continuum to the blue side continuum of 4000 $\AA$. The wavelength ranges for the D4000 measurement are $[4050,4250]$ and $[3750,3950]$ $\AA$ for the red and blue side continua, respectively \citep{bruz83}. The D4000 strength is mainly affected by the combination of several metal lines of constituent stars, including Ca {\sc ii} H and K lines \citep{hami85}. Considering the physical properties of stars such as temperature, metallicity, and surface gravity, stars whose spectral types are later than G0 are known to show the strongest D4000 features \citep{bruz83,hami85}.

Considering the specifications of the ACS/G800L instrument as low resolution spectroscopy, the grism data sets require the wider definition of D4000 instead of its narrower definition of \cite{balo99}. The D4000 strengths for our sample of galaxies are measured with the $HST$ ACS/G800L Grism data from the PEARS survey as mentioned in \ref{subsec:pears}. The reduced grism spectra from individual position angles (PA) were combined in the observed frame and we then de-redshifted the observed spectra to the rest-frame. De-redshifting was done by employing the redshift information (`$z_{\rm best}$' column) provided by the SED-fitting stellar mass catalog of \cite{sant15} that we are adopting in this work. Within the rest-frame spectra, we identified the blue and red continua and calculated the flux ratio of the two continua following the definition of the D4000 strength as described in this section. The errors of our D4000 measurements ($\delta$(D4000)) are also calculated considering the grism spectra fluxes and their measurement uncertainties within the D4000 bandpasses. The typical $\delta$(D4000) of our sample of galaxies that will be described in Section \ref{subsec:sample selection} is 0.09, and the typical relative error (i.e., $\delta$(D4000)/D4000) is 0.06. Further details on the D4000 measurements of this sample will be described by B. Joshi et al. (2018, in preparation).

\subsection{Sample Selection}
\label{subsec:sample selection}
The sample galaxies for this study are constructed by employing the four different catalogs described in Section \ref{sec:sample}. First, we cross-match between the PEARS master catalog\footnote{https://archive.stsci.edu/prepds/pears/}, the visually-classified morphology catalog \citep[i.e.,][]{kart15}, the stellar mass catalog \citep[i.e.,][]{sant15}, and the galaxy structural parameters catalog \citep[i.e.,][]{vand12} by cross-matching the right ascension and declination information with a cross-matching radius of 0.05$''$. The cross-matching results in 2923 galaxies. We further reduce the number of sample galaxies with the consideration of the appropriate wavelength coverage of the ACS/G800L instrument (that is, 5500 to 10500 $\AA$ \citep{kumm09}) for measuring the D4000 strength, which corresponds to the redshift range of 0.6 $\lesssim z \lesssim$ 1.2. This redshift range leaves 1050 galaxies out of 2923 galaxies.

Also, we consider the reliability of our D4000 measurements in our sample selection. First, we removed unphysically extreme values of D4000 measurements such as those within the range of D4000 $<$ 0.5 or D4000 $>$ 2.5. These extreme D4000 ranges are unlikely to be generated by the associated constituent stellar populations of galaxies, considering stellar population analysis \citep{bruz83,hami85,kauf03a}. The exclusion of these extreme, unphysical D4000 galaxies removes 20 galaxies out of 1050 galaxies, leaving 1030 galaxies.

Additionally, we adopt the error of D4000 measurements derived in Section \ref{subsec:d4000 measurement} in order to select galaxies with only moderate error of D4000 measurements. We apply the relative error cut of D4000 measurements to be smaller than 0.5. That is, we require $\delta(\rm{D4000}) / \rm{D4000} < 0.5$.
This relative error cut further removes 9 galaxies out of 1030 galaxies, leaving 996 galaxies.

With the remaining 996 galaxies, we derive the stellar mass completeness for our sample of galaxies following the methodology of \cite{pozz10}.  The limiting stellar mass ($M_{\rm{lim}}$) is calculated based on the apparent magnitude of galaxies and the limiting magnitude of the observation data set. Since our analysis incorporates the four different catalogs (see Section \ref{sec:sample} for details), the limiting magnitude for calculating the stellar mass completeness could in principle be complex. However, the morphology catalog of \cite{kart15} has selected their sample galaxies with $HST$ WFC3/IR $H$ band $< 24.5$. This is the most stringent cut among the input catalogs, and therefore {\it is} our limiting magnitude.   
The galaxy structural parameters and stellar mass catalogs \citep[that is,][respectively]{vand12,sant15} have fainter $H$-band limits. The PEARS survey catalog is magnitude limited at $i<26.5$, which is a substantially different wavelength, but we have checked that $H-i < 2$ would occur for a trivially small number of objects for our purposes.  

As described in \cite{pozz10} and \cite{huer16}, we first calculate the limiting stellar mass based on each object using the relation given as ${\rm{log}}(M_{\rm{lim}}/M_{\odot}) ={\rm{log}}(M_{*}/M_{\odot}) + 0.4(H - 24.5)$ where $M_{*}$ and 24.5 indicate a stellar mass of a galaxy and the limiting magnitude of the observation data set, respectively. And then, we calculate the 90th percentile of the $M_{\rm{lim}}$ distribution in each redshift bin below which the 90 $\%$ of the $M_{\rm{lim}}$ lie. In this way, the derivation ensures that the stellar mass of our sample galaxies is complete at least up to 90 $\%$.

We derive the 90th percentile of the $M_{\rm{lim}}$ distribution with only quiescent galaxies because they typically have higher mass-to-light ratios than star-forming galaxies and thus have a higher stellar mass completeness threshold than star-forming galaxies. For this derivation, we adopt the $UVJ$ color-color criterion for selecting quiescent galaxies suggested by \cite{will09} (their Equation 4) as follows:
\begin{eqnarray}
(U - V) > 0.88 \times (V - J) + 0.49. \qquad \quad [1.0 < z < 2.0]
\label{Eq1}
\end{eqnarray}
With the quiescent galaxies selected by the $UVJ$ criterion, the derived 90th percentile of the $M_{\rm lim}$ distribution within the redshift range $1.0 < z < 1.2$, which is the highest redshift range of our sample of galaxies, is ${\rm{log}}(M_{*}/M_{\odot}) = 9.44$. Our stellar mass completeness can be compared with that in \cite{huer16}, where the same limiting magnitude in the $H$ band (i.e., $H < 24.5$) and the same photometric catalog for the GOODS-South field \citep[i.e.,][]{guo13} are used. Their stellar mass completeness within the redshift range of $1.1 < z < 1.5$ is ${\rm{log}}(M_{*}/M_{\odot}) = 9.61$ for their `All' sample galaxies, which shows reasonable agreement with our derived value of ${\rm{log}}(M_{*}/M_{\odot}) = 9.44$. This stellar mass completeness of ${\rm{log}}(M_{*}/M_{\odot}) = 9.44$ reduces our sample size from 996 to 379 galaxies.

Lastly, we apply the morphology classification for our sample of galaxies as described in Section \ref{subsec:morphology}. Of 379 galaxies, 86, 61, 133, and 72 galaxies are classified as spheroids, early-type disks, late-type disks, and irregulars, respectively. The remaining 27 galaxies do not belong to any of the four morphological types and thus, are excluded from our sample of galaxies. This last sample selection criterion leaves the final sample of 352 galaxies which will be analysed throughout this paper.

\begin{figure*}[ht]
\centering
\includegraphics[width=1.0\textwidth]{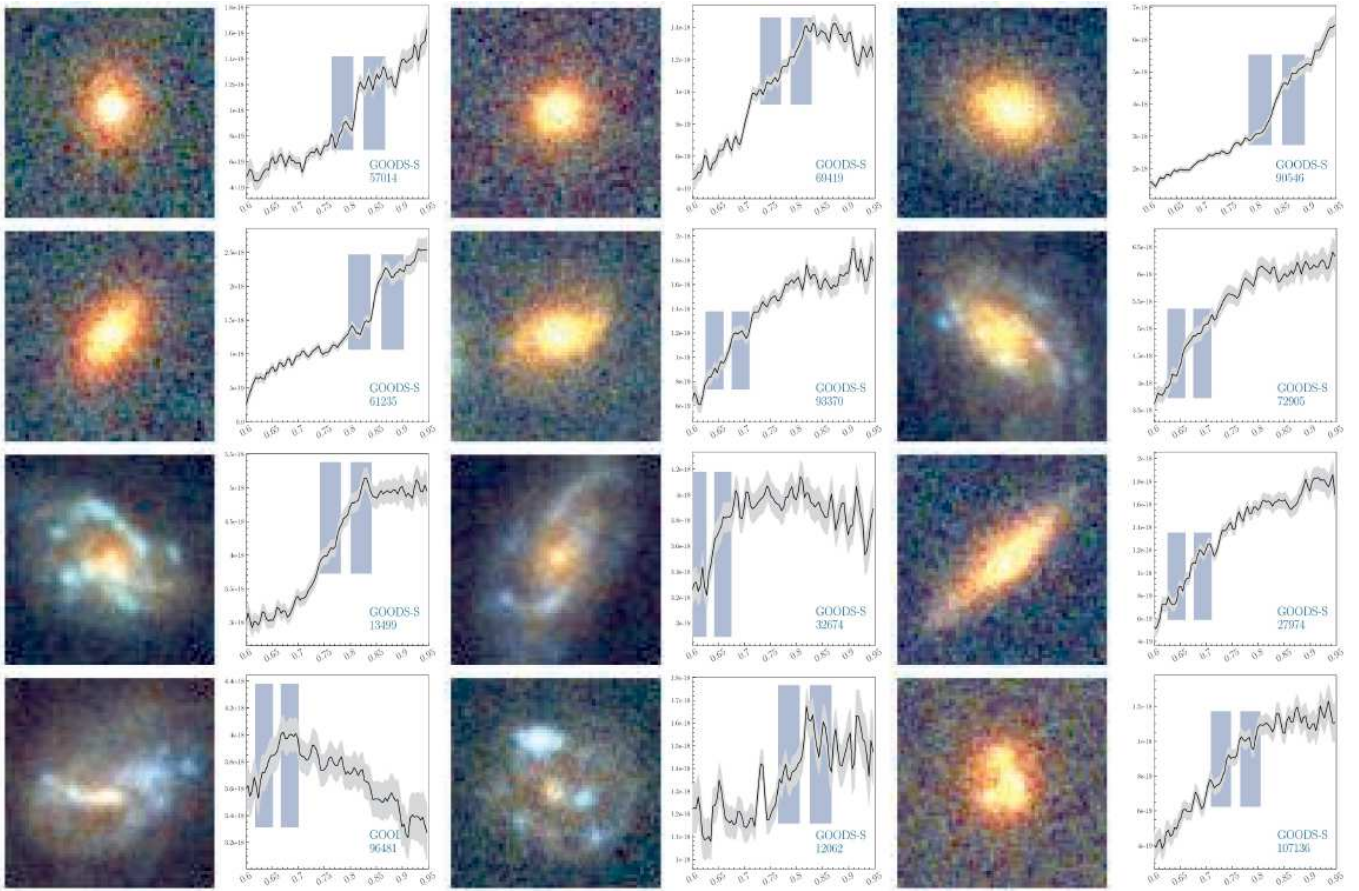}
\caption{Examples of the ACS $BVz$ filters color composite images (odd columns) and the corresponding PEARS grism spectra (even columns) of sample galaxies. All of the color composite images are 3$''\rm{x}$ 3$''$ sized. The wavelength range of the grism spectra is given in observed frame wavelength with the unit of \unit{\micro\meter}.. The unit of the flux of the grism spectra is given as erg s$^{-1}$cm$^{-2}$ \angstrom. The black solid line is the observed spectrum and the grey-shaded area is the corresponding 1$\sigma$ error. The two blue bands indicate the blue and red continua for D4000 measurement, respectively. The PEARS ID is marked on bottom right. Each row contains three randomly selected galaxies in each of the four morphological types. From top to bottom, visually classified spheroids, early-type disks, late-type disks, and irregulars are displayed. See Section \ref{subsec:morphology} for further detail on the morphology classification scheme adopted in this study.}
\label{fig1}
\end{figure*}

The number of sample galaxies in each sample selection criterion is summarized in Table \ref{tab1}. Also, examples of our sample of galaxies in each morphological type with their color-composite images and the PEARS Grism spectra are shown in Figure \ref{fig1}.

\begin{table}
\centering
\begin{threeparttable}
\caption{Summary of Sample Selection}
\begin{tabular}{ll}
\hline \hline
Criterion & Explanation (Number of galaxies)\\
\hline
Cross-matching & Cross-matching radius of\\
the four catalogs\tnote{a} & 0.05$''$ (2923)\\
\hline
0.6 $\lesssim z \lesssim$ 1.2 & Redshift range for the D4000\\
& measurement with the PEARS\\ 
&Grism data (1050)\\
\hline
0.5 $<$ D4000 $<$ 2.5 $\&$ & Remove the poor measurements\\
$\delta$(D4000)/D4000 $<$ 0.5\tnote{b} & of D4000 (996)\\
\hline
log$(M_{*}/M_{\odot}) > 9.44$ & Stellar mass completeness (379)\\
\hline
Morphology classification& Remove 27 galaxies whose\\
-Spheroids (Sph) & morphology classification\\
-Early-type Disks (E-D) & does not belong to any\\
-Late-type Disks (L-D) & of the four morphological types of\\
-Irregulars (Irr) &  Sph, E-D, L-D, and Irr\\
 & (86, 61, 133, and 72, respectively)\\
\hline
Total & 352\\
\hline \hline \\
\label{tab1}
\end{tabular}
{\small
\begin{tablenotes}
\item[a] The four catalogs used in this analysis. See Section \ref{sec:sample} for details.
\item[b] The error of D4000 measurements derived as described in Section \ref{subsec:d4000 measurement}.
\end{tablenotes}
}
\end{threeparttable}
\end{table}

\subsection{Comparison of Visually-Classified Morphology with Other Morphology Indicators}
\label{subsec:visual morphology vs CAS}
In this section, we compare the visually-classified morphology of our sample of galaxies with other widely-used non-parametric morphology indicators such as concentration index, asymmetry, and $M_{\rm 20}$ \citep{bers00,cons03,lotz04}. For the comparison, we employ the concentration index, asymmetry, and $M_{\rm 20}$ values measured by \cite{ferr05,ferr09a} based on the $HST$/ACS $i$ band. 62 galaxies among our final sample of 352 morphologically classified galaxies have been measured for concentration index, asymmetry, and $M_{\rm 20}$ values from \cite{ferr05,ferr09a}.

The concentration index we employ here is defined as the ratio of the radius enclosing 80 \% of the flux to that enclosing 20 \% of the flux. The asymmetry is calculated using the squared-pixel values of residual images after an image of the galaxy is rotated by 180$^\circ$ about its center and subtracted from the original (unrotated) image.  The asymmetry is applied with noise correction as well (see \cite{cons03} for further details). $M_{\rm 20}$ is defined as the second-order moment of the brightest 20 $\%$ of the galaxy's flux, normalized by the total second-order moment of the galaxy's flux \citep{lotz04}.

\begin{figure}
\centering
\includegraphics[width=0.5\textwidth]{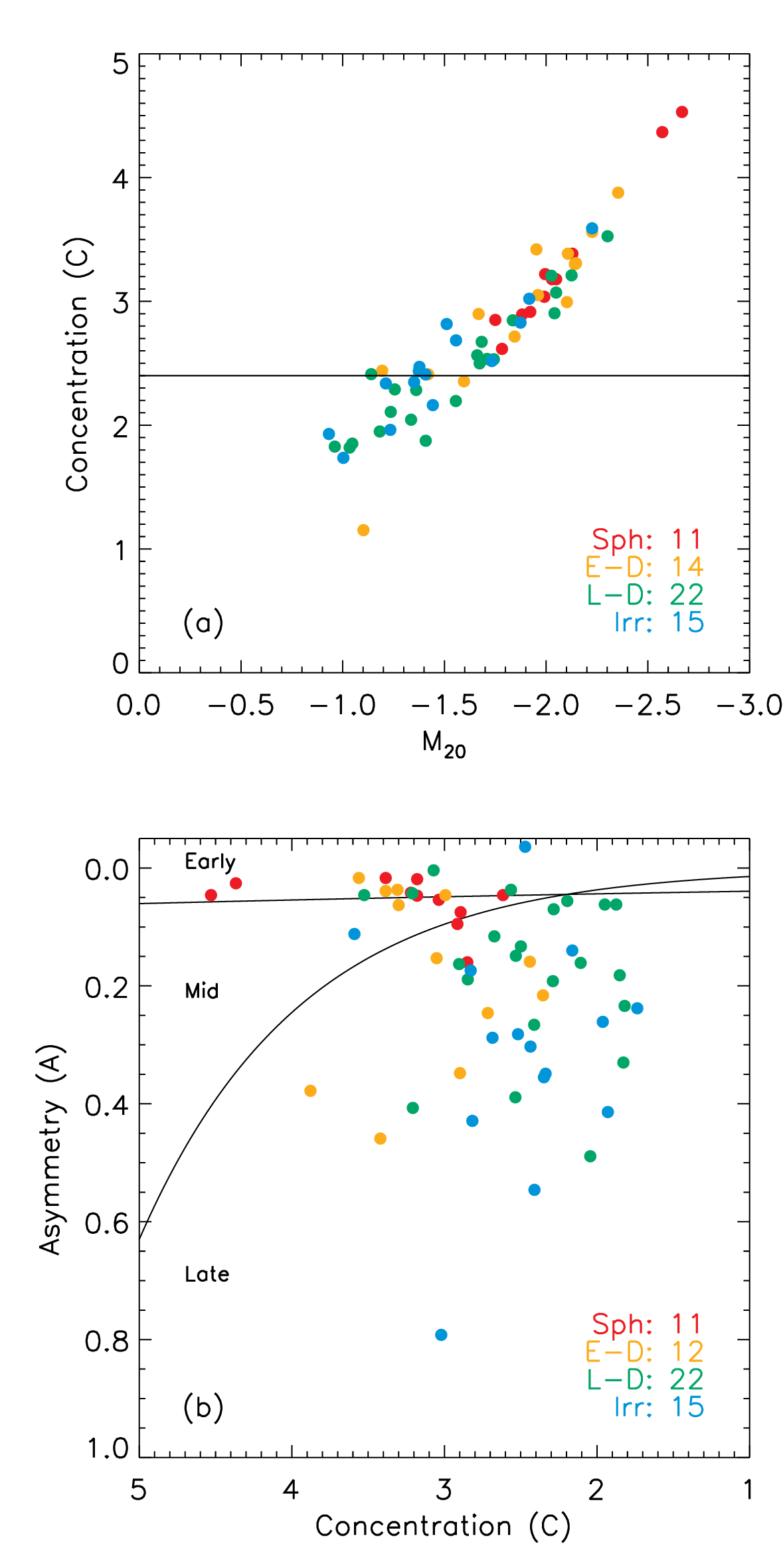}
\caption{Comparison of visually classified morphology with other morphology indicators. (a) Concentration vs. $M_{20}$ \citep*{lotz04} for four morphological types color-coded. The morphological types and the number of galaxies in each type are marked in bottom right corner. The black solid horizontal line is the suggested line for selecting early-type candidates from \cite{ferr05}. (b) Asymmetry vs. Concentration for the same sample of galaxies as panel (a). The two black solid lines are the suggested lines for separating early, mid, and late-type galaxies from \cite{bers00}. The format is the same as panel (a).}
\label{fig2}
\end{figure}

Figure \ref{fig2} shows the correlations between the concentration index, asymmetry, $M_{\rm 20}$, and the visual morphology classification. Panel (a) shows the concentration index versus $M_{\rm 20}$, with the visual-inspection morphology color-coded. The black horizontal line of concentration index of 2.4 is the criterion that \cite{ferr05} adopted for selecting their early-type candidates. The overall distribution of galaxies in panel (a) shows a positive correlation between the two morphology indicators of concentration index and $M_{\rm 20}$, as shown in previous studies \citep{lotz04,ferr05}.

Figure~\ref{fig2} additionally shows how well the visual morphology we employ in this paper is correlated with concentration index, asymmetry, and $M_{\rm 20}$. The morphology distribution in panel (a) seems consistent with both concentration index and $M_{\rm 20}$ distributions, such that all spheroids and $\sim$ 86 $\%$ of early-type disks (that is, 12 out of 14 galaxies) are located above the black horizontal line of concentration index of 2.4 along the positive correlation. The distributions of late-type disks and irregulars seem broader than those of spheroids and early-type disks, ranging from low concentration index of $\sim$ 1.6 and high $M_{\rm 20}$ of $\sim$ -1.0 to larger concentration index and lower $M_{\rm 20}$ values similar to those of spheroids and early-type disks. These broader distributions of late-type disks and irregulars in concentration index vs. $M_{\rm 20}$ seem to originate from their more complex galaxy structure (as compared to spheroids and early-type disks), which may include variously sized bulge components, diverse shapes of spiral arms, and so on.

Panel (b) shows the correlation between asymmetry and concentration index with the visual morphology. The two black solid lines are the demarcation lines for classifying galaxies into early, mid, and late types as suggested in \cite{bers00}. 2 of 14 early-type disks are removed in this panel because one of the two has an extremely large value of asymmetry (i.e., 1.194), and the other has a poorly measured asymmetry. Most spheroids (i.e., 9 of 11 galaxies) are located in the early or mid types regions having relatively small asymmetry values. However, early-type disks, late-type disks, and irregulars show broader distributions in the panel. Visual inspection of these early-type disks, late-type disks, and irregulars confirms that late-type disks and irregulars with larger asymmetry values (approximately larger than 0.3) tend to show more significantly disturbed features than those with smaller asymmetry values.

Overall, the visual morphology classification employed in this paper shows reasonable agreement with other widely-used morphology indicators such as concentration index, asymmetry, and $M_{\rm 20}$ as explored in Figure \ref{fig2}.

\section{Results}
\label{sec:results}

\subsection{Morphology Distribution in the UVJ diagram with the D4000 Strength}
\label{subsec:morphology in the UVJ}
There is a well-known correlation between galaxy morphology and their underlying stellar populations such that elliptical and S0 (i.e., early-type) galaxies consist of relatively old and metal rich stars, while spiral and irregular (i.e., late-type) galaxies consist of relatively young and metal poor stars based on their luminosity-weighted ages and metallicity analysis \citep{oh13,scha14,alpa15,khim15}. The correlation between galaxy morphology and stellar properties of galaxies has thus suggested that the understanding of galaxy morphology is necessary in order to fully understand galaxy formation and evolution \citep[see][for a review]{cons14}.

Both color-magnitude or color-color diagrams have been widely employed in numerous previous studies as ways to explore the stellar properties of galaxies photometrically \citep{sand78,bowe92,stra01,blan03,driv06,will09}. Among the diverse possible combinations of colors, we adopt the $UVJ$ diagram in this section.

\begin{figure*}[ht]
\centering
\includegraphics[width=1.0\textwidth]{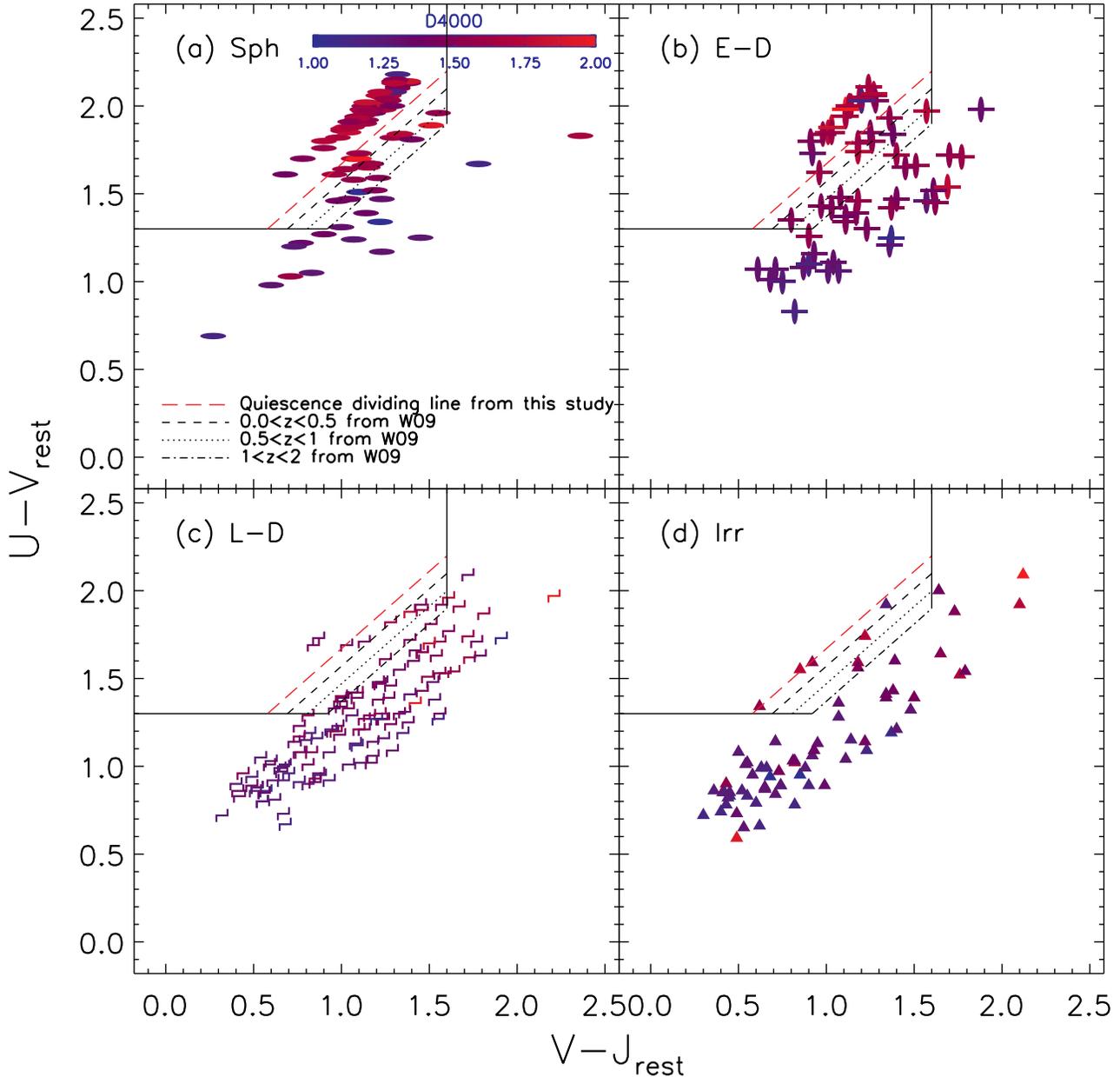}
\caption{The $UVJ$ diagram for different morphological types of galaxies with the D4000 strength color-coded. From top left to bottom right, spheroids, early-type disks, late-type disks, and irregulars are shown. The black lines are the suggested lines from \cite{will09} for separating photometrically star-forming galaxies and quiescent counterparts. The red dashed line is the suggested line from this study for identifying both photometrically and spectroscopically red sequence galaxies in the $UVJ$ diagram. Note the lack of late-type disks and irregulars above our suggested red dashed line. The overall morphological distributions of galaxies in the $UVJ$ diagram qualitatively agree well with those in \cite{kelk17} for galaxies at $0.4 \leq z \leq 0.8$ (see their Figure 6).}
\label{fig3}
\end{figure*}

Figure \ref{fig3} shows the $UVJ$ diagram for our sample of galaxies. We marked the black lines for identifying photometrically-quiescent galaxies from \cite{will09}, where the degeneracy between dust-free quiescent galaxies and dust-obscured starburst galaxies is empirically broken. Taking advantage of having both galaxy morphology and the D4000 strength information, it is interesting to note that our analysis in the $UVJ$ diagram is able to show how galaxy morphology and the photometric and spectroscopic properties of galaxies are correlated at intermediate redshifts (i.e., 0.6 $\lesssim z \lesssim$ 1.2), which is expected to shed some light on the understanding of star formation histories of galaxies along with their morphological properties.

We separately plot the $UVJ$ diagram for the four morphological subsamples described in section~\ref{subsec:morphology}, namely, spheroids, early-type disks, late-type disks, and irregulars.   We indicate their D4000 strengths in Figure \ref{fig3} using point colors. First of all, a reasonable correlation between the optical colors of $U-V$ and $V-J$ and the D4000 strength is clearly shown for all morphological types, such that galaxies with redder optical colors mostly have larger D4000 strength. This correlation overall ranges from blue to red colors of $U-V$ and $V-J$ (i.e., approximately from 0.5 to 2.2 for $U-V$ and from 0.5 to 2.0 for $V-J$).

\begin{figure*}[ht]
\centering
\includegraphics[width=1.0\textwidth]{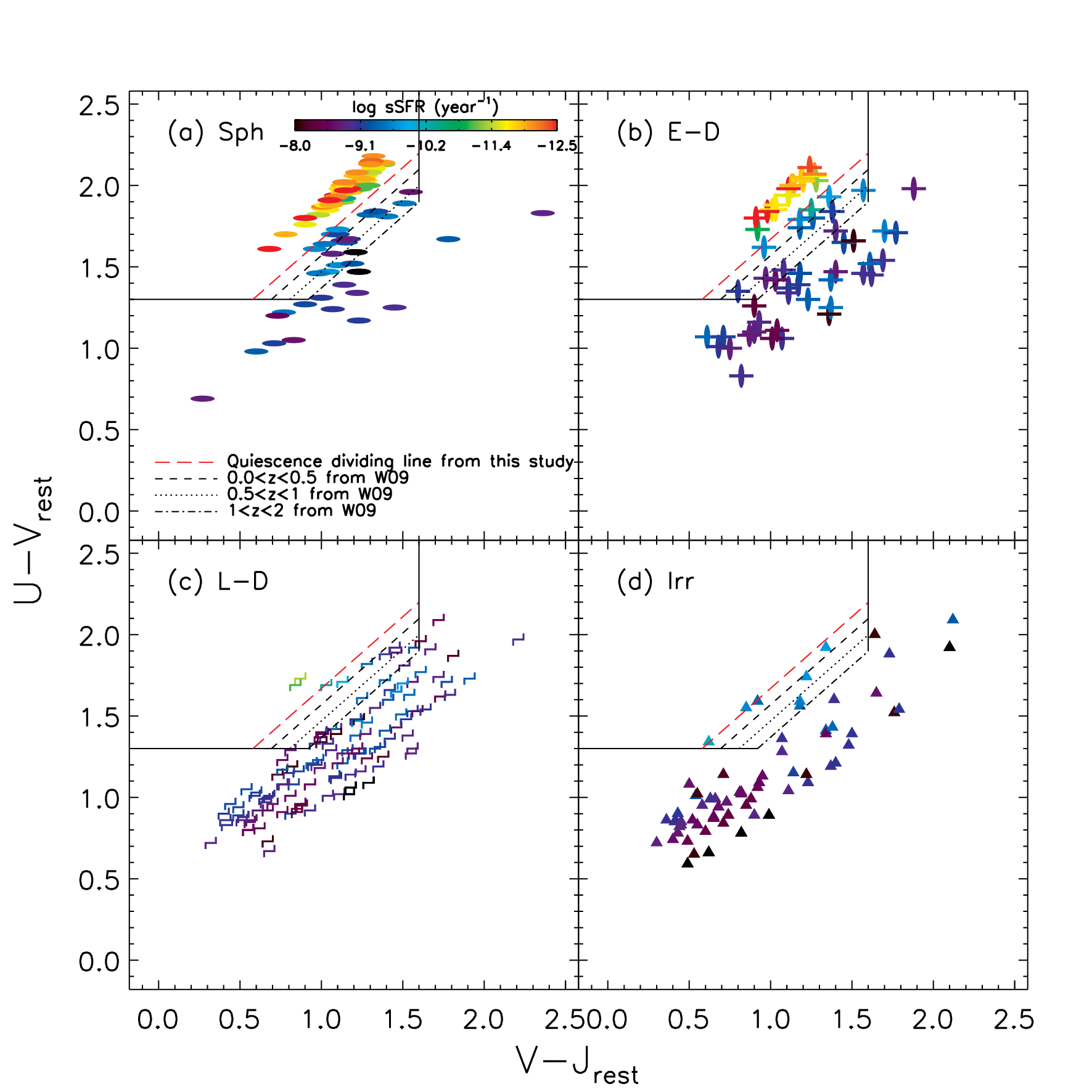}
\caption{The same $UVJ$ diagram as Figure \ref{fig3}, but with galaxy specific star formation rate (sSFR). Note that our newly suggested criterion (i.e., the red dashed line in each panel) for identifying optically quiescent galaxies based on the D4000 strength is qualitatively in good agreement with the distribution of sSFR. See the text for details.}
\label{fig3_2}
\end{figure*}

Panel (a) shows the $UVJ$ diagram for \textit{spheroids}. The majority of spheroids (i.e., 64 of 86 galaxies, 74 \%) are within the region of quiescence (i.e., top-left region) defined with the black dotted line suggested by \cite{will09} for the redshift range $0.5 < z < 1.0$, within which more than half of our sample (that is, 205 of 352 galaxies) of galaxies reside. A color sequence of quiescent spheroids with increasing both $U-V$ and $V-J$ colors in the quiescent region is noticeable. The quiescent sequence of `red and dead' spheroids is apparently so tight that it seems reasonable to suggest another separation line for defining the region of quiescence in the $UVJ$ diagram based on our sample of galaxies within the redshift range 0.6 $\lesssim$ \lowercase{$z$} $\lesssim$ 1.2. The suggested line is shown in red dashed line in each panel and is as follows:
\begin{eqnarray}
(U - V) > 0.88 \times (V - J) + 0.79. \qquad \quad [0.6 \lesssim z \lesssim 1.2]
\label{Eq2}
\end{eqnarray}
The newly suggested line has basically the same slope as that in \cite{will09}, but has a 0.1 larger intercept in order to define the quiescent region more conservatively, mainly focusing on the tight quiescent sequence as shown in panel (a). The line is found to be well-correlated with the D4000 strength such that the average D4000 strength for quiescent spheroids defined with the new line is 1.60. In addition to the D4000 strength, the color evolutionary track with an exponentially declining star formation history with $\tau = 0.1$ Gyr in the $UVJ$ diagram presented in \cite{owns16} shows a close overlapping region with the red sequence of our spheroids. Moreover, the same $UVJ$ diagram but with galaxy specific star formation rate (that is, SFR divided by stellar mass) in Figure \ref{fig3_2} shows a qualitatively good agreement with the distribution of the D4000 strength. Given the overall large D4000 strengths, the close overlapping of the color evolutionary track with the stellar population model, and the qualitatively good agreement with the specific star formation rates, our newly suggested line seems to reasonably select the optically authentic `red and dead' galaxies at intermediate redshifts based on their colors and the D4000 strength.

The sequence of quiescent spheroids is effectively the same sequence of quiescent galaxies identified in previous studies \citep{will09,whit11,bell12,muzz13,owns16}, while additionally showing their visually-identified morphological properties and spectroscopic properties represented by the D4000 strength. Considering the distributions of all other types of morphologies (i.e., \textit{early-type disks}, \textit{late-type disks}, and \textit{irregulars} in panels (b), (c), and (d), respectively) in the \textit{UVJ} diagram as well, it is interesting to note that the majority (i.e., 66 of 68 galaxies, $\sim$ 97\%) of the quiescent galaxies defined with the red dashed line are bulge-dominated systems being classified as either spheroids or early-type disks. This strong correlation between the dominance of a bulge component and the quenching of star formation activities in galaxies is qualitatively in agreement with what has been known as the correlation between galaxy structure and star formation activities \citep{wuyt11,bell12,sain12,bait17}.

However, it is also clear that not all spheroids and early-type disks (i.e., the bulge-dominated systems) have fully quenched their star formation activities, as there exist relatively blue (that is, $U-V$ $\lesssim$ 1.3 and $V-J$ $\lesssim$ 1.0) and weak D4000 strength (that is, D4000 $\lesssim$ 1.5) spheroids and early-type disks. Those of not-quenched bulge-dominated systems are distributed from the blue colors of $U-V$ ($\sim$ 0.7) and $V-J$ ($\sim$ 0.5) to the locations just below the red sequence. Along with the gradual changes in their optical colors, their D4000 strength is increasing as well. These gradually increasing trends in both the $UVJ$ colors and the D4000 strength are also seen in late-type disks and irregulars in panels (c) and (d), respectively. The gradual changes in the optical properties of galaxies seem to show the transitional sequence of galaxies from the `blue cloud' to the `red sequence' as shown in the color-magnitude (or stellar mass) relations in previous studies \citep{bald04,bell04,scha14}.

While there are photometrically and spectroscopically `red and dead' quiescent galaxies in spheroids and early-type disks,  few late-type disks and irregulars are within the region of quiescence defined with our suggested line (i.e., the red dashed lines in panels (c) and (d)). Only 2 of 133 late-type disks and none of 72 irregulars (1.5 \% and 0 \%, respectively) are within the quiescent region. Even if we consider the other quiescence lines (i.e., the black lines in the panels), the fraction of quiescent galaxies in late-type disks and irregulars is still low: 6.0 \% of late-type disks and 6.9 \% of irregulars would be classified as quiescent using the black dotted line, compared with 74\% of spheroids and 43\% of early-type disks.  Although there are gradually increasing trends in the $UVJ$ colors along with the D4000 strength from the `blue cloud' to the `red sequence' in late-type disks and irregulars, the lack of these bulgeless systems within the region of quiescence in the $UVJ$ diagram seems to show again the close correlation between the galaxy structure and the quenching of star formation activities in galaxies, as mentioned earlier in this section.

\subsection{The D4000 strength vs. Stellar Mass and Stellar Surface Density with Morphologies}
\label{subsec:D4000 vs. stellar mass and surface density}
The D4000 strength of galaxies (or similarly, its narrower wavelength range version D$_{\rm n}$4000;  \citet{balo99}) is a useful luminosity-weighted stellar population age indicator, as explained in Section \ref{subsec:d4000 measurement}. In particular, a D4000 strength of 1.5 generally corresponds to an average stellar population age of 1 Gyr, and thus the D4000 strength of 1.5 or a similar value has been used as a criterion for separating star-forming galaxies and quiescent galaxies \citep{kauf03a,kauf03b,hath09,hain17}.

We explore the correlations between the D4000 strength, stellar mass, the S\'ersic index, and stellar surface density within the half-light radius for different subsets of our final sample, to check if these stellar property-related correlations show any dependence on morphological types.

\begin{figure*}[ht]
\centering
\includegraphics[width=1.0\textwidth]{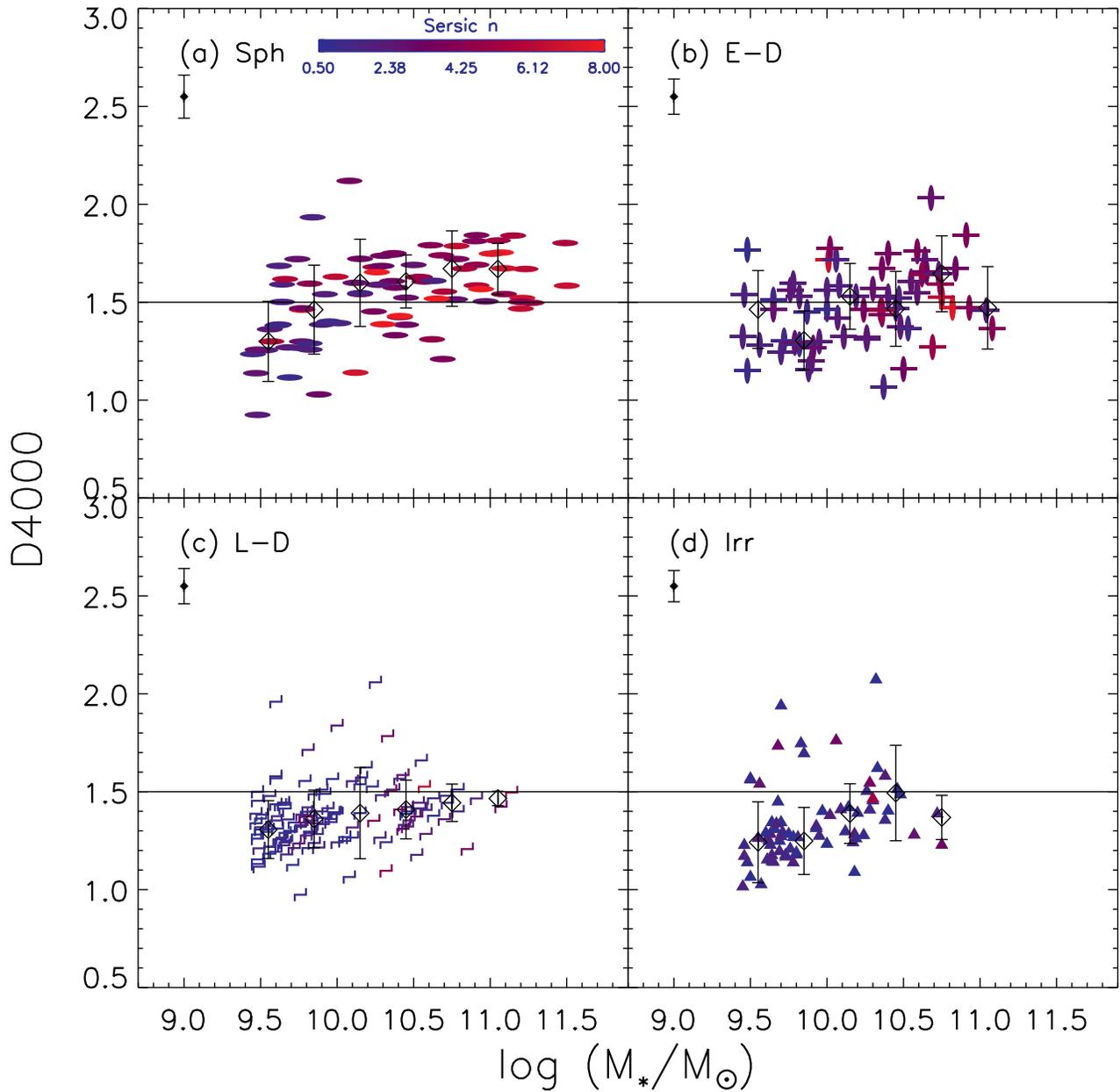}
\caption{The D4000 strength vs. stellar mass with the S\'ersic index color-coded. The format is the same as in Figure \ref{fig3}. The black-solid horizontal line is the D4000 strength of 1.5 for distinguishing spectroscopically star-forming galaxies (i.e., D4000 $<$ 1.5) and quiescent galaxies (i.e., D4000 $\geq$ 1.5). The empty diamonds are the running median values and the corresponding error bars show the 1$\sigma$ standard deviations to indicate the population scatter. The typical measurement uncertainty in the D4000 strength for each morphological type is marked on top left of each panel.}
\label{fig4}
\end{figure*}

Figure \ref{fig4} shows the correlation between the D4000 strength and stellar mass for galaxies of each morphological type. The color indicates the S\'ersic index, and the empty diamonds and the corresponding error bars indicate the median values and 1-$\sigma$ standard deviations of D4000 strength distribution in each stellar mass bin, respectively. Panel (a) shows the distribution for spheroids. There is a moderately positive correlation between the D4000 strength and stellar mass, based on the overall distribution and the median values marked in the panel. The Pearson correlation coefficient (the associated \textit{p}-value) is only 0.48 (0.00), although the value is slightly larger than the same correlation coefficient of 0.44 (0.00) for the total sample of galaxies, which is most likely attributed to the quality of the grism spectra and the vulnerability of the D4000 index to the possible outshining (i.e., luminosity-weighted ``frosting'') effect of small amount of young stellar populations in the majority of relatively old stellar populations (see \cite{trag00,hern13} for further discussion on this). This increasing trend has similarly (but, with respect to the entire population of galaxies without morphology classification) been shown with the D$_{\rm n}$4000 within similar redshift range in previous studies and has been discussed as one of the indicators for the $``$downsizing$"$ scenario for the star formation histories of galaxies \citep{cowi96,delu06,hern13,paci16,hain17}. Despite some overlapping of the 1$\sigma$ standard deviations of the median values near the D4000 strength of 1.5, all of the median values for stellar mass bins greater than log$(M_{*}/M_{\odot}) \gtrsim 10.0$ show the D4000 strengths greater than 1.5, suggesting that most of the moderately massive spheroids (i.e., log$(M_{*}/M_{\odot}) \gtrsim 10.0$) observed at intermediate redshift (i.e., 0.6 $\lesssim$ \lowercase{$z$} $\lesssim$ 1.2) ceased forming stars at least 1 Gyr prior to the observed epoch. However, there are also some fraction of spheroids that show D4000 $< 1.5$. Those weak-D4000 spheroids tend to be less massive (i.e., log$(M_{*}/M_{\odot}) \lesssim 10.0$). The fraction of strong D4000 galaxies in each mass bin for the four morphological types is summarized in Table \ref{tab2}. As in the table, 62.8 $\%$ of our sample of spheroids show strong D4000 strength.

The S\'ersic index for each plotted galaxy is indicated by color in Figure \ref{fig4}, allowing examination of its dependence on D4000 strength and stellar mass. The distribution shows that massive (thus, likely large D4000 strength) spheroids tend to have larger values of S\'ersic index, suggesting that massive spheroids have either centrally concentrated or extended light profiles, or both. We will discuss the S\'ersic index distribution and the concentration of light profiles of galaxies in more details in Section \ref{subsec:the concentration of galaxy light profiles}.

The moderately positive correlations between the D4000 strength, stellar mass, and the S\'ersic index shown in spheroids seem gradually weaker as the morphology of interest changes from spheroids to late-type disks. Panels (b) and (c) show the correlations between the quantities in early-type disks and late-type disks, respectively. Early-type disks show 47.5 $\%$ of strong D4000 galaxies among them for the entire stellar mass range, which is 0.76 times that of spheroids (i.e., 62.8 $\%$). However, the positive correlation between the D4000 strength and stellar mass is still visible based on the running median values (the empty diamonds) in panel (b), although the increasing trend is not as strong as in spheroids and the associated Pearson correlation coefficient (the associated \textit{p}-value) is only 0.30 (0.02). Late-type disks in panel (c) show an interestingly and relatively flat trend between the D4000 strength and stellar mass, even up to a higher stellar mass regime, log$(M_{*}/M_{\odot}) \sim 11.0$. The Pearson correlation coefficient (the associated \textit{p}-value) for late-type disks is 0.25 (0.00). The running median values and the fraction of strong D4000 galaxies among late-type disks do not seem to show any strong indication of the positive correlation between the D4000 strength and stellar mass that is visible in spheroids. Only 16.5 $\%$ of late-type disks within the entire stellar mass range of interest have strong D4000 strength, which is 0.26 times that of spheroids.

Moreover, among the 16.5 $\%$ of late-type disks with strong D4000 (that is, 22 late-type disks as shown in Table \ref{tab2}), many may not actually be quenched. 15 of these 22 objects (i.e., 68 $\%$) have an apparent axis ratio less than 0.5, suggesting that the strong D4000 strength in late-type disks is mainly caused by their inclination and the associated extinction. For instance, using the D4000 index with the stellar population synthesis models of \cite{vazd12} at solar metallicity, the effect of an E(B-V) = 0.2 mag attenuation results in a dust-free equivalent age change of 0.5 Gyr for a 1 Gyr old population, or an age change of 2 Gyr for a 3Gyr old population. Additionally, a test on the effect of E(B-V) on the measured value of the D4000 strength was performed. Specifically, we considered an intrinsic D4000 strength of 1.5 and applied the extinction law from \cite{calz00}. For the case of color excess E(B-V) of 0.2 mag attenuation, the intrinsic value of 1.5 was changed to 1.603, which is $\sim$ 7 $\%$ increase. For the case of color excess E(B-V) of 0.4 mag attenuation, the change is from 1.5 to 1.714, $\sim$ 14 $\%$ increase, and for the case of color excess E(B-V) of 0.6 mag attenuation, the change is from 1.5 to 1.832, $\sim$ 22 $\%$ increase. Therefore, the inclination effect on late-type disks seems to further reduce the fraction of authentically `red and dead' late-type disks, corroborating that late-type disks do not get quenched and do not show a strong mass quenching at intermediate redshifts (i.e., 0.6 $\lesssim$ \lowercase{$z$} $\lesssim$ 1.2).

Panel (d) shows the correlations in irregulars. The Pearson correlation coefficient (the associated \textit{p}-value) for irregulars is 0.31 (0.01). They show qualitatively similar trends that are shown in late-type disks in low stellar mass regime (that is, log$(M_{*}/M_{\odot}) \lesssim 10.0$) such that most of low stellar mass irregulars (that is, approximately 83 $\%$ of the less massive irregulars based on Table \ref{tab2}) do not show strong D4000 strengths. These low stellar mass irregulars mainly show clumpy and ill-defined features such as asymmetric disks and non-spherical components at their centers based on their color-composite images, which is expected for them to be classified as irregulars. However, in the higher stellar mass regime (that is, log$(M_{*}/M_{\odot}) \gtrsim 10.0$), the running median values of D4000 strength increase as stellar mass increases. This positive correlation between the D4000 strength and stellar mass is likely to be attributed to irregulars that have irregular features mainly due to galaxy interactions such as mergers and flyby, rather than typically clumpy star-forming irregulars. Visual inspection on the color-composite images of moderately massive irregulars (i.e., stellar mass larger than log$(M_{*}/M_{\odot}) > 10.1$) shows that more than half of these massive irregulars have more than one of the visually-identifiable systems, suggestive of the likely interactions with other galaxies. Having the two distinctive types of irregulars in our morphology classification is presumably due to the fact that the classification scheme adopted in our study for classifying irregulars does not distinguish between the typically clumpy star-forming irregulars and the galaxy interaction-driven irregulars (see Section \ref{subsec:morphology} for details).

Regarding the correlation between the D4000 strength and stellar mass, we additionally calculate the Kendall's Tau nonparametric correlation coefficient \citep{ken38}. The results of the correlation coefficient (the associated \textit{p}-value) are 0.33 (0.00), 0.34 (0.00), 0.21 (0.02), 0.23 (0.00), and 0.31 (0.00) for the total sample galaxies, spheroids, early-type disks, late-type disks, and irregulars, respectively, which suggests that the overall trend shown in the Pearson correlation coefficients for the correlation is consistent with that shown in the Kendall's Tau nonparametric correlation coefficients.

The statistical significance on the Pearson correlation coefficients and the Kendall's Tau nonparametric correlation coefficients for the correlation between the D4000 strength and stellar mass is further checked by performing a bootstrap resampling in addition to the associated \textit{p}-values. That is, we performed 1000 bootstrap resampling by shuffling the stellar masses while fixing the D4000 strengths within each morphological type and measured the correlation coefficients accordingly. From the 1000 bootstrap resampling results, the number of occurrences that the Pearson correlation coefficient from each bootstrap resampling is larger than the actually calculated value are 0, 0, 13, 2, and 4 for the total sample, spheroids, early-type disks, late-type disks, and irregulars, respectively. In case of the Kendall's Tau nonparametric correlation coefficient, the number of occurrences are 0, 0, 6, 0, and 0 in the same order of morphological type as the one for the Pearson correlation coefficient. In both correlation coefficients, the results from the bootstrap resampling test show the qualitatively similar trend to that shown in the associated \textit{p}-values to the actually measured correlation coefficients and seem to suggest that our computed correlation coefficients are not likely randomly obtained.

Another useful statistics for analysing the distributions of parameters of interest, the Cram\'er-von Mises statistics \citep[][]{cram28,vonm28} for the two-dimensional distribution of two samples \citep[][]{ande62} with bootstrap resamples (hereafter, C-vM), is also performed for the correlation between the D4000 strength and stellar mass of two selected morphological types, which could show a morphological dependence on the ``downsizing'' trend. The hypothesis tested is whether ``x is distributed as y'' where x and y indicate two selected morphological types in this case. The results of the statistical test show that the hypothesis for all pairs of the two selected morphological types is rejected with the associated \textit{p}-values of 0. The only exception when the hypothesis is confirmed is the case between spheroids and early-type disks with the associated \textit{p}-value of 0.16, most likely originated from their close similarity of morphology. We will discuss the results of the statistical test in detail in Section \ref{subsec:morphological dependnce on the downsizing trend}.

\begin{table*}[ht]
\centering
\begin{threeparttable}
\caption{Fraction of strong D4000 galaxies in each morphological type (D4000 $\geq 1.5$)}
\begin{tabular}{lllll}
\hline \hline
Mass range and Morphology (Number of galaxies) & Spheroids & Early-type Disks & Late-type Disks & Irregulars\\
\hline
9.44 $\leq$ log$(M_{*}/M_{\odot}) < 9.9$ & 28 $\%$ (7/25) & 37.5 $\%$ (6/16) & 7.4 $\%$ (5/68) & 17.1 $\%$ (7/41)\\
9.9 $\leq$ log$(M_{*}/M_{\odot}) < 10.3$ & 68.8 $\%$ (11/16) & 40 $\%$ (6/15) & 30 $\%$ (9/30) & 15.8 $\%$ (2/19)\\
10.3 $\leq$ log$(M_{*}/M_{\odot}) < 10.7$ & 66.7 $\%$ (14/21) & 57.1 $\%$ (12/21) & 26.9 $\%$ (7/26) & 40 $\%$ (3/10)\\
10.7 $\leq$ log$(M_{*}/M_{\odot}) < 11.1$ & 100 $\%$ (16/16) & 55.6 $\%$ (5/9) & 12.5 $\%$ (1/8) & 0 $\%$ (0/2)\\
11.1 $\leq$ log$(M_{*}/M_{\odot}) \leq 11.5$ & 75 $\%$ (6/8) & 0 galaxy & 0 $\%$ (0/1) & 0 galaxy\\
\hline
Total & 62.8 $\%$ (54/86) & 47.5 $\%$ (29/61) & 16.5 $\%$ (22/133) & 16.7 $\%$ (12/72)\\
\hline \hline \\
\label{tab2}
\end{tabular}
\end{threeparttable}
\end{table*}

It is also interesting to examine the correlations between the D4000 strength and stellar surface density as a function of morphology, since stellar surface density within the effective radius ($\Sigma_{\rm e}$) or the central 1 kpc has been suggested to be associated with the quenching of star formation activities in galaxies \citep{bell00,kauf03b,fran08,bell12,fang13,oman14,barr15,willi17,lee18}.

\begin{figure*}[ht]
\centering
\includegraphics[width=1.0\textwidth]{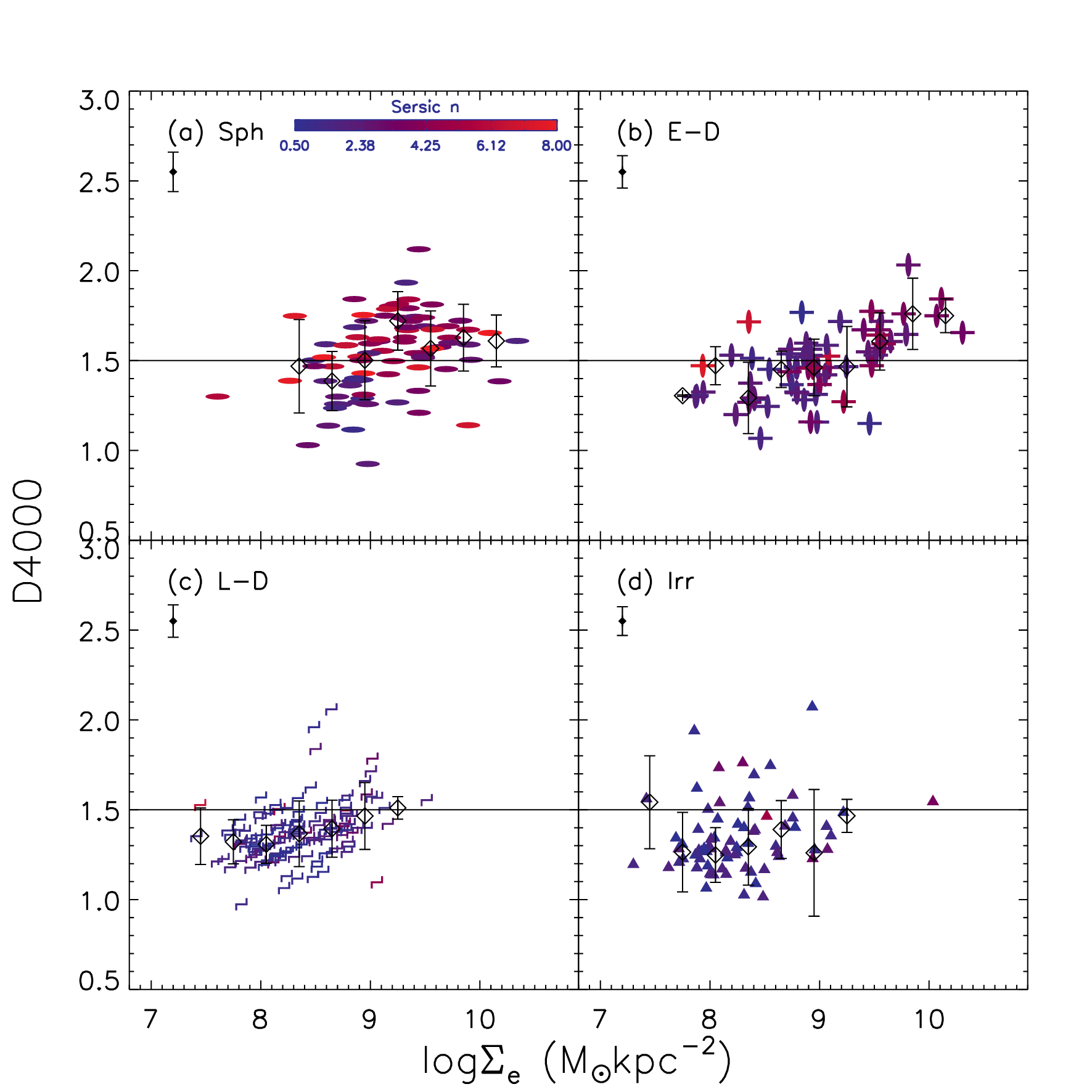}
\caption{The D4000 strength vs. stellar surface density within the half-light radius $\Sigma_{\rm e}$, with the S\'ersic index color-coded. The format is the same as Figure \ref{fig4}. Note the lack of late-type disks and irregulars with the stellar surface density larger than $10^{9}$ $M_{\odot}\rm{kpc^{-2}}$. See the text for details.}
\label{fig5}
\end{figure*}

Figure \ref{fig5} shows the correlations between the D4000 strength, $\Sigma_{\rm e}$, and the S\'ersic index that is color-coded in each data point, with separate panels for distinct morphological classes. $\Sigma_{\rm e}$ is derived adopting the widely employed formula: $\Sigma_{\rm e} = 0.5M_{*}/(\pi r_{\rm e}^{2})$ where $M_{*}$ is galaxy stellar mass in unit of solar mass, and $r_{\rm e}$ is the effective radius of galaxies in unit of kpc \citep{kauf03b,fran08,bell12}. We employ the circular effective radius for calculating the $\Sigma_{\rm e}$ and the Kormendy relation in Section \ref{subsec:kormendy relation} by multiplying the semi-major axis effective radius by the square root of the axis ratio of galaxies.

Panel (a) shows the correlations between the parameters in spheroids. The surface density distribution approximately ranges from $10^{8.3}$ $M_{\odot}\rm{kpc^{-2}}$ to $10^{10.2}$ $M_{\odot}\rm{kpc^{-2}}$. A weakly positive correlation between the D4000 strength and the surface density is seen, based on the distribution and the running median values, although the median values do not dramatically increase with increasing surface density. We compute for this correlation a Pearson correlation coefficient and the resulting correlation coefficient (the associated \textit{p}-value) is 0.24 (0.03). It is particularly interesting to compare the trend shown in the figure with the well-known sharp threshold in $\Sigma_{\rm e}$ over which galaxies suddenly become quiescent of their star formation activities \citep{brin04,fran08,whit17,willi17}. A threshold of $\Sigma_{\rm e}$ which varies depending on the redshift of interest roughly corresponds to $10^{9}$ $M_{\odot}\rm{kpc^{-2}}$ at intermediate redshift (that is, the redshift range of interest in this study) \citep{whit17}. Indeed, our spheroids show a larger fraction of strong D4000 galaxies (D4000 $\geq 1.5$) above the $\Sigma_{\rm e}$ of $10^{9}$ $M_{\odot}\rm{kpc^{-2}}$, which is 80 \% (39 out of 49 galaxies), compared to 41 \% (15 out of 37 galaxies) for those below the $\Sigma_{\rm e}$ of $10^{9}$ $M_{\odot}\rm{kpc^{-2}}$. The dramatic increase in the fraction of strong D4000 galaxies above the $\Sigma_{\rm e}$ threshold is also seen in early-type disks in panel (b) with the fractions of 76 \% (19 out of 25 galaxies) and 28 \% (10 out of 36 galaxies) for galaxies above and below the $\Sigma_{\rm e}$ threshold, respectively. The Pearson correlation coefficient (the associated \textit{p}-value) for early-type disks in panel (b) is 0.58 (0.00). For reference, the Pearson correlation coefficient (the associated \textit{p}-value) for the total sample galaxies is 0.49 (0.00).

Therefore, our data support the hypothesis that $\Sigma_{\rm e}$ (or, similarly, the stellar surface density in the central 1 kpc) is related to the quenching of star formation activities in galaxies as previously reported \citep{fran08,oman14}, suggesting the role of internal structure of galaxies in quenching of star formation activities.

What is shown in late-type disks and irregulars (i.e., panels (c) and (d), respectively) is that the two types of morphologies have very few galaxies whose $\Sigma_{\rm e}$ is larger than $10^{9}$ $M_{\odot}\rm{kpc^{-2}}$, compared to spheroids and early-type disks. That is, only 8 of 133 late-type disks (6 \%) and 5 of 72 irregulars (7 \%) have the $\Sigma_{\rm e}$ larger than $10^{9}$ $M_{\odot}\rm{kpc^{-2}}$, while spheroids and early-type disks show the fractions of 49 of 86 galaxies (57 \%) and 25 of 61 galaxies (41 \%), respectively. The running median D4000 values in late-type disks do not show noticeably significant increase with increasing $\Sigma_{\rm e}$, although the associated Pearson correlation coefficient (the associated \textit{p}-value) is 0.37 (0.00). The running median D4000 values in irregulars seem rather randomly distributed with some increases and decreases as the associated Pearson correlation coefficient (the associated \textit{p}-value) of 0.22 (0.07) indicates. The lack of disk-dominated galaxies and irregular galaxies with $\Sigma_{\rm e} \gtrsim 10^{9}$ $M_{\odot}\rm{kpc^{-2}}$ is most likely to be attributed to the star-forming disk or clumpy structures in late-type disks and irregulars that typically result in a larger effective radius and the subsequently smaller $\Sigma_{\rm e}$ at given stellar mass, compared to spheroids and early-type disks.

As performed for the correlation between the D4000 strength and stellar mass, we calculate the same Kendall's Tau nonparametric correlation coefficient for the correlation between the D4000 strength and $\Sigma_{\rm e}$. The results of the correlation coefficient (the associated \textit{p}-value) are 0.37 (0.00), 0.21 (0.00), 0.42 (0.00), 0.28 (0.00), and 0.15 (0.06) for the total sample galaxies, spheroids, early-type disks, late-type disks, and irregulars, respectively. The results suggest that  the overall trend shown in the Pearson correlation coefficients for the correlation is consistent with that shown in the Kendall's Tau nonparametric correlation coefficient.

The same bootstrap resampling test as the one performed for the correlation between the D4000 strength and stellar mass to check statistical significance is performed for the correlation between the D4000 strength and $\Sigma_{\rm e}$ by shuffling $\Sigma_{\rm e}$ while fixing the D4000 strength within each morphological type. The results for the Pearson correlation coefficient are 0, 4, 0, 0, and 39 for the total sample, spheroids, early-type disks, late-type disks, and irregulars, respectively. The results for the Kendall's Tau nonparametric correlation coefficient are 0, 1, 0, 0, and 36 in the same order of morphological type as the one for the Pearson correlation coefficient. As in the correlation between the D4000 strength and stellar mass, the results from the bootstrap resampling test for the correlation between the D4000 strength and $\Sigma_{\rm e}$ also seem to suggest that our computed correlation coefficients are not likely randomly obtained.

The same C-vM test as the one performed for the correlation between the D4000 strength and stellar mass is also performed for the correlation between the D4000 strength and $\Sigma_{\rm e}$. The results show that the hypothesis ``x is distributed as y'' is rejected for all pairs of the two selected morphological types, with the associated \textit{p}-values of 0.02 for the case between spheroids and early-type disks and 0 for the other cases.

The S\'ersic index distribution that is color-coded in each data point seems to behave as expected such that higher surface density galaxies tend to have higher values of the S\'ersic index \citep{bell12}. Further analysis of the surface density and the S\'ersic index will be discussed in Section \ref{subsec:the concentration of galaxy light profiles}.

\subsection{sSFR vs. Stellar Mass and Stellar Surface Density with Morphologies}
\label{subsec:sSFR vs. stellar mass and stellar surface density}
The diagram of galaxy specific star formation rate (sSFR) versus stellar mass has often been employed to analyse the properties of stellar populations of the star-forming blue cloud, the quiescent red sequence, and the transient green valley galaxies \citep{noes07,ciam13,abra14,bren15,bren17,eale17,pand17}. Along with this diagram, the correlation between sSFR and the $\Sigma_{\rm e}$ also has been investigated to find some link between star formation activities and internal structure of galaxies \citep{whit15,whit17,willi17}.  (Similar relations using SFR in place of sSFR, and/or central surface density in place of effective-radius surface density, have similar applications.)  In this section, we explore the sSFR versus stellar mass and the surface density within the effective radius with galaxy morphology in order to see if there is any morphological dependence on the locations of galaxies in the two diagrams.

\begin{figure*}[ht]
\centering
\includegraphics[width=1.0\textwidth]{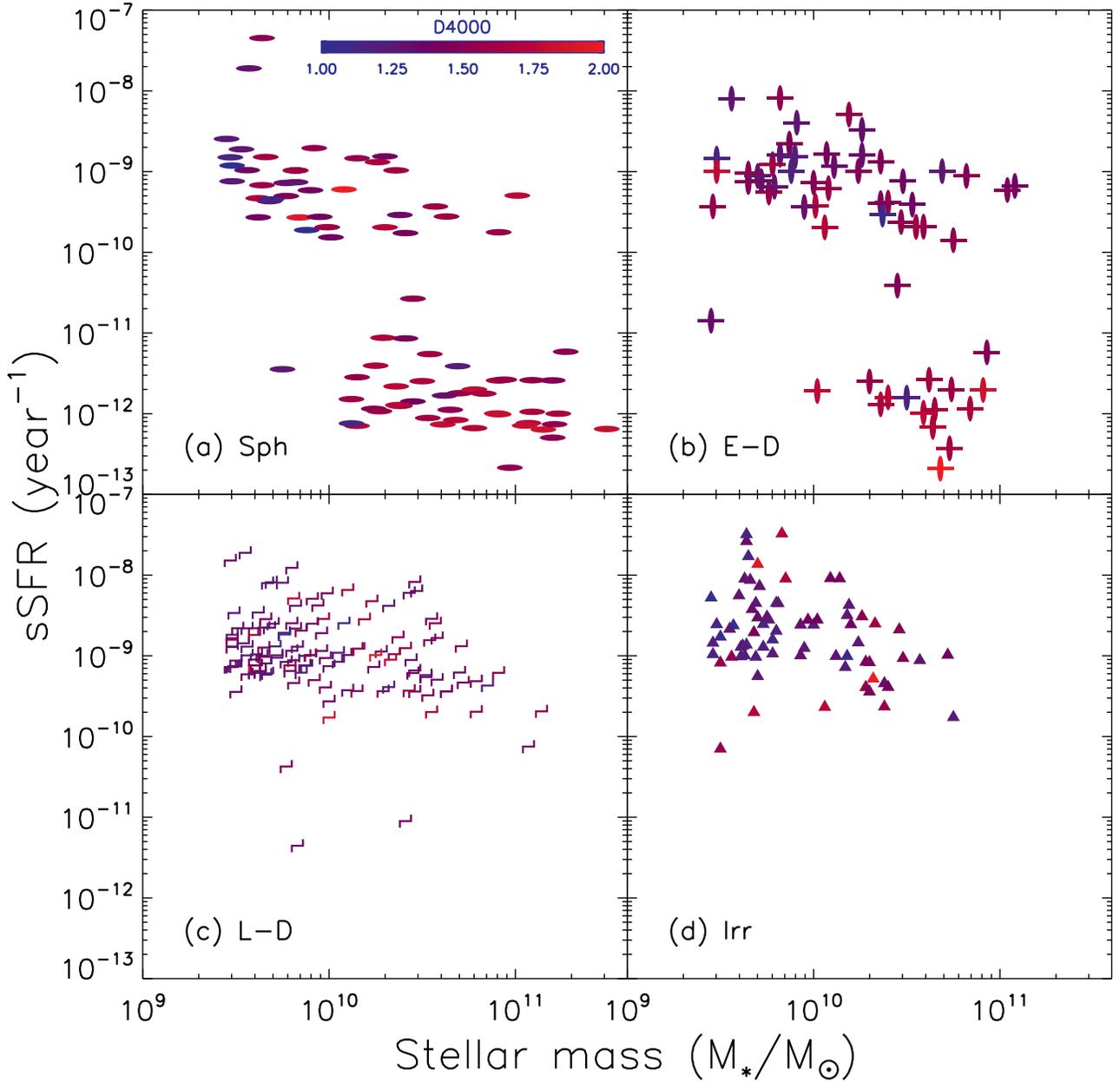}
\caption{sSFR vs. stellar mass with the D4000 strength color-coded. The format is the same as Figure \ref{fig3}. Note that galaxies with a prominent bulge component (i.e., spheroids and early-type disks) are separated into two regions depending on their sSFR values, while most of late-type disks and irregulars do not show such remarkable separations. See the text for detail.}
\label{fig6}
\end{figure*}

Figure \ref{fig6} shows the correlation between the sSFR and stellar mass for our sample of galaxies depending on their morphologies. The D4000 strength is color-coded in each data point. Panel (a) shows the correlation in spheroids. First of all, there is a clear bimodality between the star-forming spheroids and the quiescent spheroids with the separation sSFR of $\sim$ $10^{-10} \rm year^{-1}$. A similar trend is seen in early-type disks in panel (b) as well. If we take the sSFR of $10^{-10} \rm year^{-1}$ as an approximate separation criterion between the star-forming population and the quiescent population, the fractions of quiescent spheroids and early-type disks are 55.8 \% (48 of 86 galaxies) and 31.1 \% (19 of 61 galaxies), respectively. These fractions are quantitatively in good agreement with those (i.e., 62.8 \% and 47.5 \%, respectively) derived by the fractions of strong D4000 galaxies in Table \ref{tab2} with slightly higher fractions in the D4000 case. The similar fractions suggest the overall consistency between the photometrically-derived sSFR and the spectroscopically-derived D4000 strength. Other than the clear bimodality, the mass quenching trend is shown in bulge-dominated systems, such that the more massive galaxies are, the lower their sSFR is. Additionally, the presence of both star-forming bulge-dominated systems (i.e., either spheroids or early-type disks) and their quiescent counterparts in panels (a) and (b) (i.e., galaxies with sSFR $\gtrsim 10^{-10} \rm year^{-1}$ and sSFR $\lesssim 10^{-10} \rm year^{-1}$, respectively) is confirmed again, as previously shown in Figures \ref{fig3}, \ref{fig3_2}, \ref{fig4}, and \ref{fig5}. The distribution of the bimodality shown in Figure \ref{fig6} is qualitatively consistent with those of \cite{sali08} and \cite{sant09}.

While spheroids and early-type disks show the overall mass quenching trend as well as the bimodality in the diagram of sSFR versus stellar mass, the distributions of late-type disks and irregulars (panels (b) and (c) in Figure \ref{fig6}, respectively) do not show a strong mass quenching trend or bimodality even up to the stellar mass of $\sim$ $10^{11} M_{*}/M_{\odot}$. In other words, there are only 4 of 133 (3.0 \%) and 1 of 72 (1.4 \%) quiescent (i.e., sSFR $\lesssim 10^{-10} \rm year^{-1}$) late-type disks and irregulars, respectively throughout the stellar mass range considered in the figure (that is, stellar mass greater than $10^{9.44}$ $M_{*}/M_{\odot}$). The remarkable lack of quiescent late-type disks and irregulars at intermediate redshift seems to confirm disk components or clumpy structures as the main drivers of global star formation activities in galaxies \citep{marg18}. This is further supported by the fact that the moderately massive (i.e., stellar mass greater than $10^{10}$ $M_{*}/M_{\odot}$) star-forming (sSFR $\gtrsim 10^{-10} \rm year^{-1}$) early-type disks still tend to show visually-identifiable blue star-forming disk components, while their quiescent counterparts (i.e., sSFR $\lesssim 10^{-10} \rm year^{-1}$) show almost spheroid-like morphologies without any visually-identifiable blue star-forming disk components. We checked the $BVz$ filters color composite images of these early-type disks and noticed the morphological differences between the star-forming early-type disks and the quiescent early-type disks even within the same massive stellar mass range (that is, stellar mass greater than $10^{10}$ $M_{*}/M_{\odot}$).

The results from the C-vM test for the correlation between sSFR and stellar mass indicate that the hypothesis ``x is distributed as y'' is rejected for all pairs of the two selected morphological types, with the associated \textit{p}-values of 0.01 for the case between spheroids and early-type disks, 0.02 for the case between late-type disks and irregulars, and 0 for the other cases.

\begin{figure*}[ht]
\centering
\includegraphics[width=1.0\textwidth]{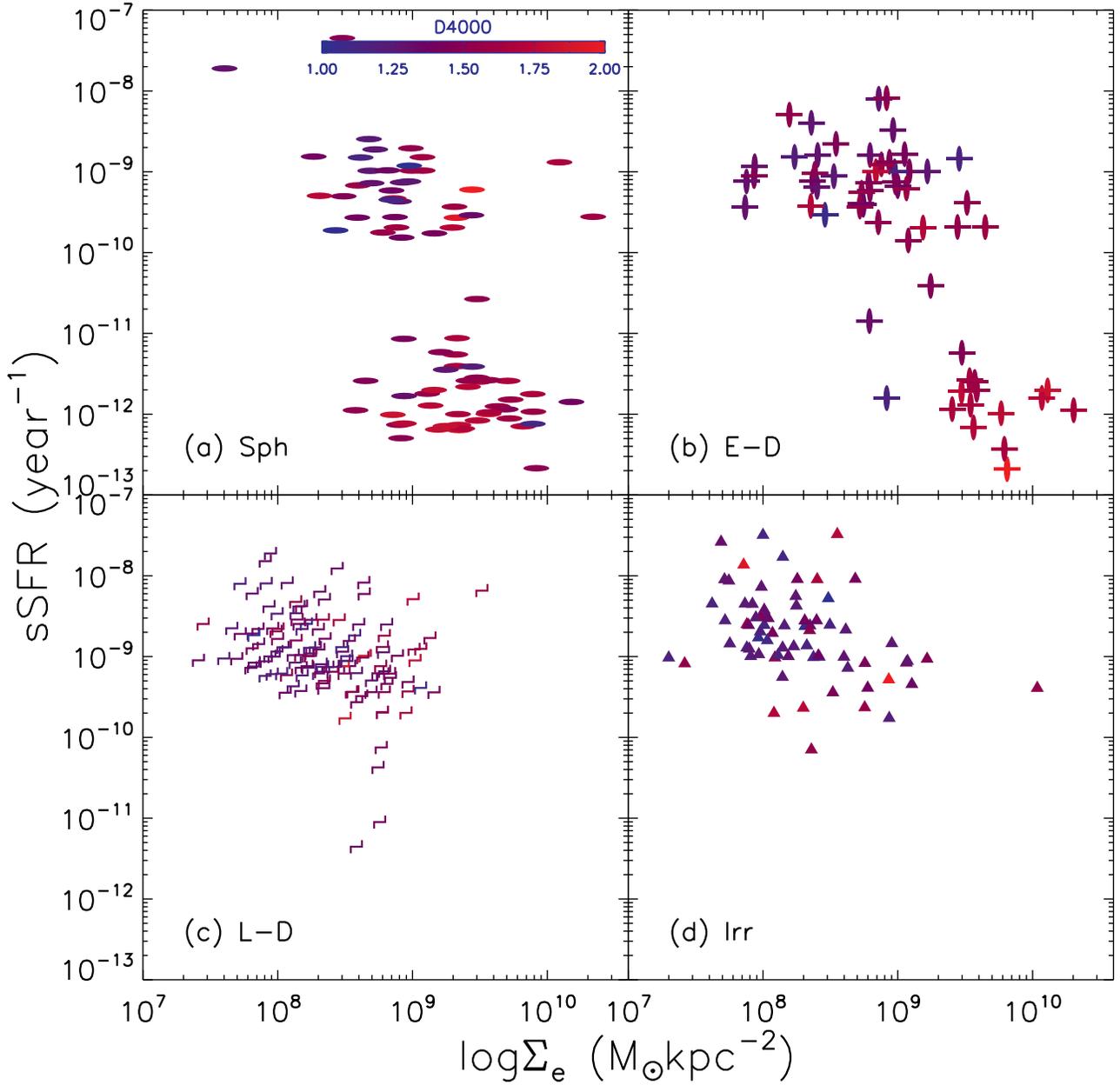}
\caption{sSFR vs. $\Sigma_{\rm e}$ with the D4000 strength color-coded. The format is the same as in Figure \ref{fig3}. Note that most of the low sSFR galaxies (i.e., sSFR $< 10^{-10}\rm{year^{-1}}$) are bulge-dominated galaxies and have the stellar surface density approximately greater than $10^{9}M_{\odot}\rm{kpc^{-2}}$, while the high sSFR galaxies (sSFR $\gtrsim 10^{-10}\rm{year^{-1}}$) in any morphological types mostly have the stellar surface density smaller than $ 10^{9}M_{\odot}\rm{kpc^{-2}}$.}
\label{fig7}
\end{figure*}

We also explore the morphological dependence on the star formation activities of galaxies in the same diagram, but for the surface density instead of stellar mass. Figure \ref{fig7} shows the correlations between sSFR and $\Sigma_{\rm e}$ with the D4000 strength color-coded. As inferred from Figures \ref{fig5} and \ref{fig6}, spheroids and early-type disks in panels (a) and (b), respectively show the bimodality such that there are star-forming and low $\Sigma_{\rm e}$ populations (i.e., sSFR $\gtrsim 10^{-10} \rm year^{-1}$ and $\Sigma_{\rm e} \lesssim 10^{9}$ $M_{\odot}\rm{kpc^{-2}}$) and quiescent and high $\Sigma_{\rm e}$ populations (i.e., sSFR $\lesssim 10^{-10} \rm year^{-1}$ and $\Sigma_{\rm e} \gtrsim 10^{9}$ $M_{\odot}\rm{kpc^{-2}}$). The separation between the star-forming galaxies and the quiescent galaxies at $\Sigma_{\rm e} \sim 10^{9}$ $M_{\odot}\rm{kpc^{-2}}$ seems clear \citep[e.g.,][]{whit17}, as discussed earlier in Section \ref{subsec:D4000 vs. stellar mass and surface density}.

The distributions for late-type disks and irregulars in panels (c) and (d), respectively show that they do not have quiescent and high $\Sigma_{\rm e}$ galaxies (i.e., sSFR $\lesssim 10^{-10} \rm year^{-1}$ and $\Sigma_{\rm e} \gtrsim 10^{9}$ $M_{\odot}\rm{kpc^{-2}}$), unlike spheroids and early-type disks. The lack of quiescent and high $\Sigma_{\rm e}$ galaxies in late-type disks and irregulars seems to be attributed to star-forming disk components (or clumpy structures for irregulars) and the associated extendedly-measured effective radii that result in small $\Sigma_{\rm e}$ in the late-type disks and irregulars.

The results from the C-vM test for the correlation between sSFR and $\Sigma_{\rm e}$ indicate that the hypothesis ``x is distributed as y'' is rejected for all pairs of the two selected morphological types, with the associated \textit{p}-values of 0.01 for the cases between spheroids and early-type disks and between late-type disks and irregulars and 0 for the other cases.

At this point, we would like to remind that the overall sSFR distributions shown in Figures \ref{fig6} and \ref{fig7} are based on the sSFR derived by the multi-wavelnegth photometric bands SED fitting procedures, as mentioned earlier in Section \ref{subsec:stellar mass}. Regarding this, we note that the SED fitting-based SFR and the associated sSFR typically show larger uncertainties and have lower nominal values than the IR and UV flux-based SFR and the associated sSFR as previously discussed \cite[][see their Figures 2 and 7]{sant09}.

Finally, we tested whether our conclusions depend on which stellar mass prescription we adopt among the several possibilities provided by \cite{sant15}. We find that the trends shown in Figures \ref{fig6} and \ref{fig7} are robust to changes in stellar mass fitting procedure, although the bimodality at sSFR $\sim 10^{-10} \rm year^{-1}$ shown in those figures can be less distinct for some of the other stellar mass models.

\subsection{The concentration of light profiles of galaxies with Morphologies}
\label{subsec:the concentration of galaxy light profiles}
The concentration of galaxies' light profiles may be related to their star formation activity.  Galaxies with highly concentrated light profiles, as probed by the S\'ersic index or central surface density, generally show the signatures of quenched star formation \citep{bell12,cheu12,barr17}. This demonstrates the predictive power of the central surface density for quiescence \citep{whit17,lee18}.

\begin{figure*}[ht]
\centering
\includegraphics[width=1.0\textwidth]{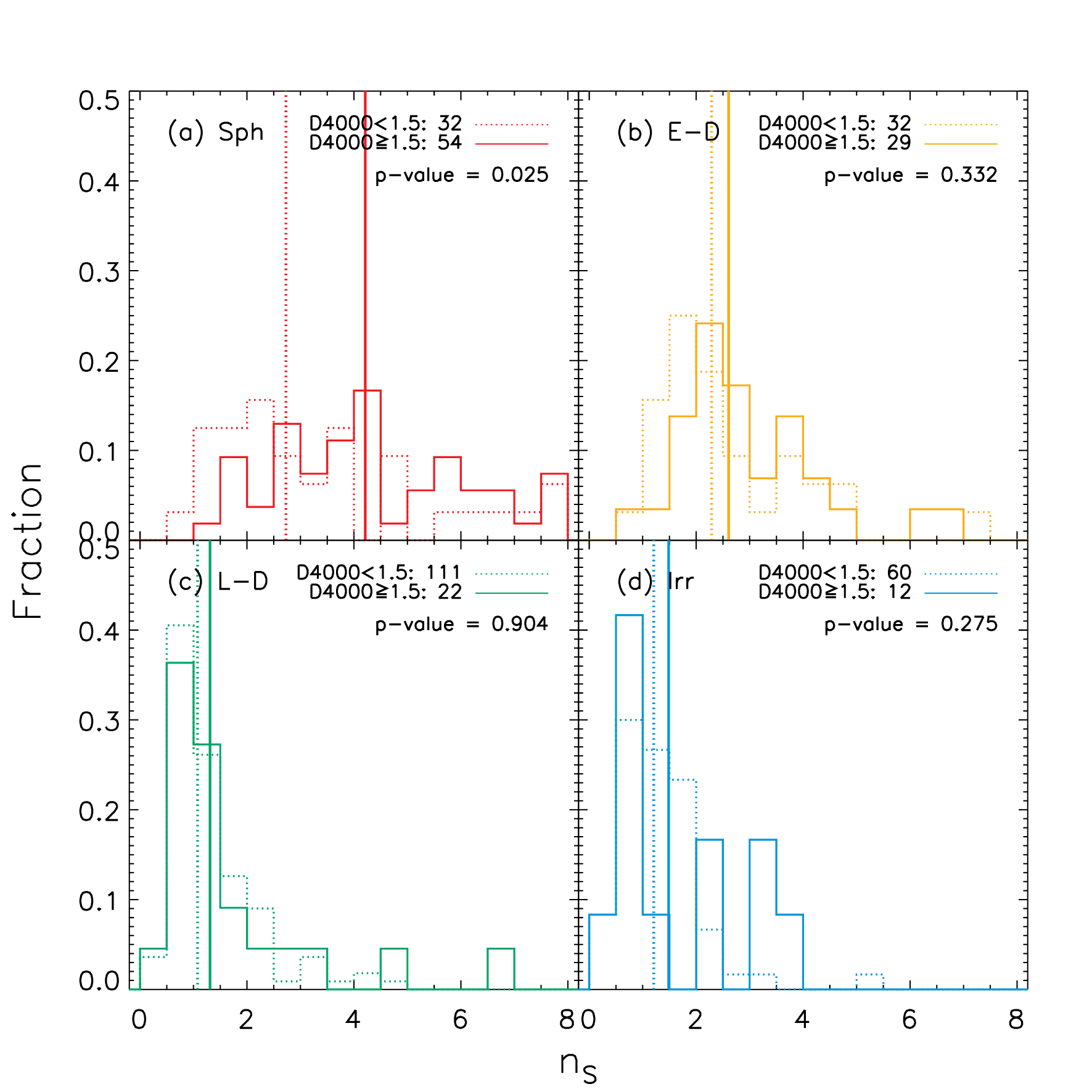}
\caption{S\'ersic index distribution for each of the four morphological types. From top left to bottom right, spheroids, early-type disks, late-type disks, and irregulars are displayed. In each panel, the dotted and solid lines correspond to the S\'ersic index distribution for weak D4000 (i.e., D4000 $< 1.5$) and strong D4000 (i.e., D4000 $\geq 1.5$) galaxies, respectively. The corresponding vertical lines indicate the median values of each D4000 strength group. The number of galaxies and the $p$-values from the K$-$S test are marked on top right of each panel. Note that the difference in distribution between the weak and strong D4000 strength groups is most significant in spheroids. The median value is n = 4.21 for strong D4000 galaxies, while the median value for the weak D4000 counterparts is only n = 2.73. However, a significant difference between weak and strong D4000 groups is not seen in other morphological types. See the text for details.}
\label{fig8}
\end{figure*}

We explore a morphological dependence of the correlation between light profile concentration and star formation quiescence using the S\'ersic index, $\Sigma_{\rm e}$, and the D4000 strength of galaxies. Figure \ref{fig8} shows the S\'ersic index distributions for the four types of morphologies adopted in this paper. In each morphological type, we divide galaxies into star-forming and quiescent galaxies depending on their D4000 strength with the same criterion of D4000 of 1.5 as employed in Table \ref{tab2}. Panel (a) shows the distribution for spheroids. 
We find that the star-forming spheroids tend to have lower S\'ersic indices (the median value is 2.73), while the quiescent spheroids tend to have higher S\'ersic indices (the median value is 4.21). Moreover, the difference is reasonably  robust. A Kolmogorov$-$Smirnov (K$-$S) test comparing the S\'ersic index distributions for the weak and strong D4000 distributions yields only a 2.5 \% probability that such a large difference would be observed from intrinsically identical distributions.

While the difference in the S\'ersic index distribution between the weak and strong D4000 spheroids seems significant, the difference becomes gradually weaker as the morphological type of interest goes from spheroids to late-type disks. The median S\'ersic index values for weak and strong D4000 early-type disks are 2.29 and 2.61, respectively. The associated K-S test $p$-value is 0.332; i.e., there is a 33.2 $\%$ probability of observing a difference this large among two samples drawn from the same intrinsic distribution. For late-type disks, the median S\'ersic indices for the weak and strong D4000 subsets are 1.08 and 1.31, a difference that is not statistically significant (with K-S test $p$-value of 0.904). Finally, irregulars have median S\'ersic index values 1.21 and 1.49, for weak and strong D4000 subsets, with an associated K-S test $p$-value of 0.275. While this difference could be due to random chance, it is also consistent with the observation that the ``irregular'' galaxies are a mixed set including galaxies of different nature, as discussed earlier in Section \ref{subsec:D4000 vs. stellar mass and surface density} (i.e., star-forming clumpy irregulars and galaxy interaction-driven irregulars).

\begin{figure*}[ht]
\centering
\includegraphics[width=1.0\textwidth]{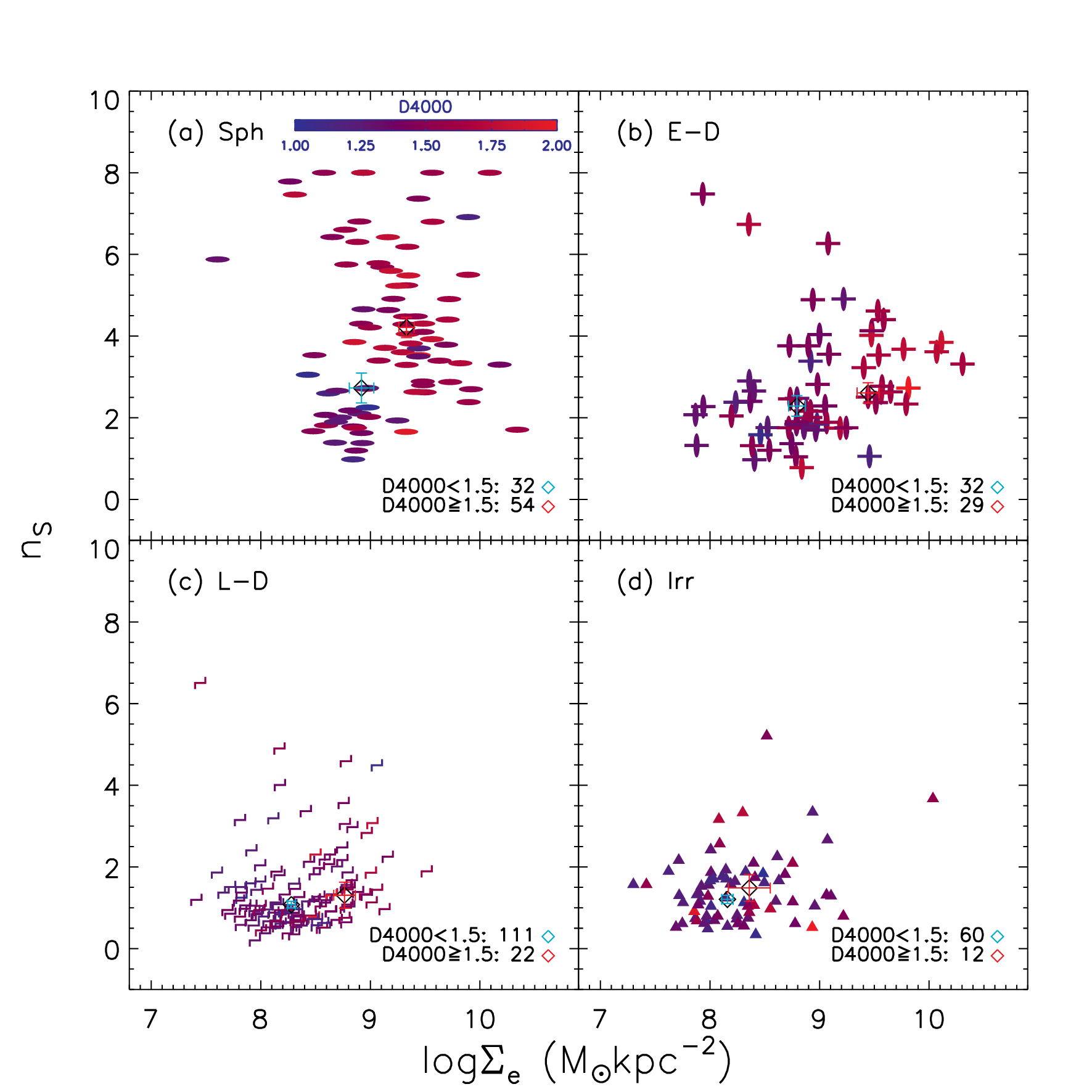}
\caption{S\'ersic index vs. stellar surface density within the half-light radius, with the D4000 strength color-coded. The format is the same as Figure \ref{fig4}. The two diamonds with blue and red error bars in each panel are the median values for weak and strong D4000 galaxies, respectively. The corresponding error bars indicate the errors of the means in each D4000 group of galaxies.}
\label{fig9}
\end{figure*}

The results from the S\'ersic index distribution for each of morphological types are further confirmed when galaxies are plotted in the diagram of the S\'ersic index versus $\Sigma_{\rm e}$, which is shown in Figure \ref{fig9} with the D4000 strength color-coded. In Figure \ref{fig9}, each morphological type is divided into the same weak and strong D4000 galaxies as in Figure \ref{fig8}. The median S\'ersic index and $\Sigma_{\rm e}$ values of each of weak and strong D4000 galaxies are also marked with empty diamonds with blue and red error bars indicating the errors of the means in each D4000 group of galaxies, respectively. Spheroids in panel (a) show that not only the strong D4000 spheroids have typically higher S\'ersic indices, but also they have larger $\Sigma_{\rm e}$ values, suggesting that the strong D4000 spheroids have more compact light profiles than the weak D4000 counterparts. A similar trend is seen in early-type disks but with a smaller difference in the S\'ersic index distribution between the weak and strong D4000 early-type disks, compared to spheroids.

The noticeable correlation between the D4000 strength and compactness explored by the S\'ersic index and $\Sigma_{\rm e}$ in spheroids and early-type disks does not seem to exist in late-type disks and irregulars, partly as expected from the analysis of Figure \ref{fig8} for the S\'ersic index distribution. Along with negligible differences in the S\'ersic index distribution between weak and strong D4000 galaxies in late-type disks and irregulars (that is, the median S\'ersic index differences of 0.23 and 0.28 between the weak and strong D4000 groups in late-type disks and irregulars, respectively), their $\Sigma_{\rm e}$ values show marginal differences between weak and strong D4000 galaxies as well. That is, late-type disks and irregulars have the median $\log(\Sigma_{\rm e})$ value differences of 0.48 and 0.2, respectively, suggesting a marginal correlation between compactness and quenching in late-type disks and irregulars.

Overall, what is shown in Figures \ref{fig8} and \ref{fig9} is a morphological dependence on the observed correlations such that bulge-dominated systems (that is, spheroids and early-type disks) have noticeably positive correlation, while disk or clumpy structure-dominated systems do not, implying that galaxies need to experience morphological transformations to bulge-dominated systems for them to follow the observed positive correlation between quenching and compactness of galaxies as shown in Figures \ref{fig8} and \ref{fig9}.

\subsection{The Kormendy Relation for Spheorids}
\label{subsec:kormendy relation}
In this section, we explore the Kormendy relation for our sample of spheroids. This scaling relation has been widely employed in order to study structural properties of early-type galaxies and spheroidal components of late-type galaxies and the associated formation and evolution mechanisms \citep{korm77,bern03b,laba03,ferr05,long07,gado09,fish10,kim16}.

\begin{figure}
\centering
\includegraphics[width=0.5\textwidth]{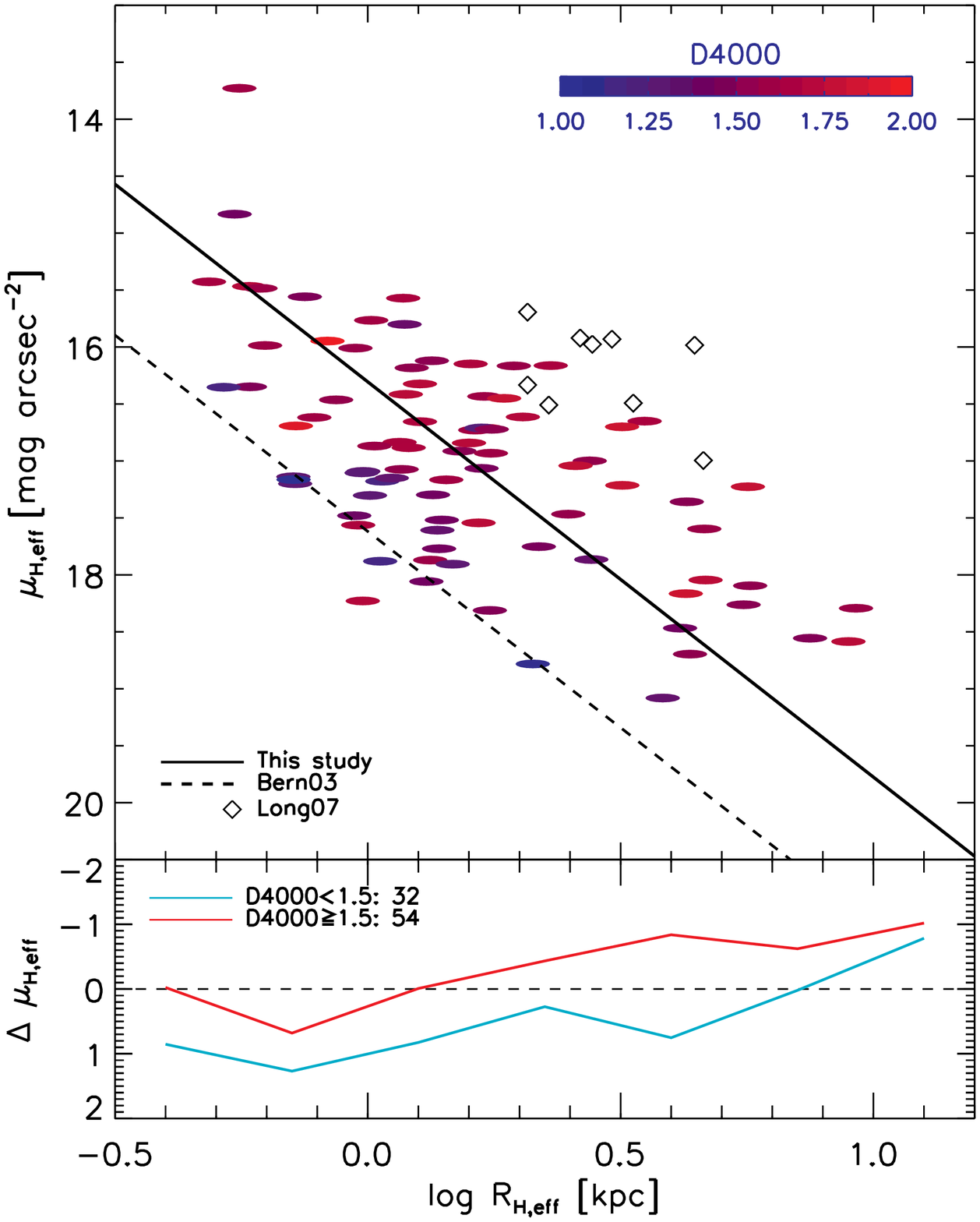}
\caption{Top panel: the Kormendy relation for spheroids, with the D4000 strength color-coded. The color bar in top right indicates the D4000 strength. The black straight line is the linear fit to our sample of spheroids. The black dashed line is the Kormendy relation for relatively local (i.e., $0.01 < z <0.3$) early-type galaxies derived in the SDSS $i$-band from \cite[Bern03b]{bern03b}. The empty diamonds are early-type galaxies at $1.2 < z < 1.7$ from \cite[Long07]{long07}. All of the data points are corrected for the cosmological dimming $(1 + z)^{4}$. Bottom panel: the fit residuals of spheroids to the Kormendy relation depending on their D4000 strength. Note that the weak D4000 strength spheroids tend to lie below the fitting line, while the strong D4000 strength spheroids tend to lie above the fitting line. See text for details.}
\label{fig10}
\end{figure}

The top panel in Figure \ref{fig10} shows the Kormendy relation for our sample of spheroids with the D4000 strength color-coded. We adopted the same circular effective radius calculated in Section \ref{subsec:D4000 vs. stellar mass and surface density} and the same mean surface brightness as defined in \cite{bern03a}. Also, we corrected for the cosmological dimming $(1 + z)^{4}$ to our sample of spheroids and 9 early-type galaxies at $1.2 < z < 1.7$ that are shown as the empty diamonds in the panel from \cite{long07}. Our spheroids seem to follow the general scaling relation based on the comparison to the 9 early-type galaxies and the fit through the maximum lilkehood analysis (the black dashed line in the panel) to $\sim$ 9,000 early-type galaxies at $0.01 < z < 0.3$ in the Sloan Digital Sky Survey (SDSS) from \cite{bern03b}. We particularly refer to the fit in the SDSS $i$-band because the bandpass of the $i$-band at low redshifts closely traces the bandpass of the $H$-band at intermediate redshifts for our sample of spheroids (that is, the median redshift of our sample spheroids is 0.94). We performed a linear fit to our sample of spheroids and derived the following fitting relation:
\begin{eqnarray}
\mu_{\rm H,eff} = (3.47 \pm 0.31) \rm{log_{10}} \textit{R}_{H,eff} + 16.31 \pm 0.11.
\label{Eq3}
\end{eqnarray}
The slope of 3.47 $\pm$ 0.31 derived for our sample spheroids shows a comparable range with that of 3.29 $\pm$ 0.09 derived for the local early-type galaxies from \cite{bern03b} except for a larger uncertainty of the fit and a slightly steeper slope value. 

Other than the general trend of the Kormendy relation shown in our sample of spheroids, we further check any dependence on the stellar properties of spheroids presented by the D4000 strength in this scaling relation. We divide the spheroids into weak and strong D4000 groups as in Section \ref{subsec:the concentration of galaxy light profiles} and calculate the residuals of the surface brightness to the linear fit that is expressed in Equation \ref{Eq3}. The residuals are shown in the bottom panel of Figure \ref{fig10}. As likely inferred from the top panel, at fixed effective radius the weak D4000 spheroids tend to have larger (fainter) surface brightness (luminosity) than the strong D4000 spheroids by $\sim$ 0.8 mag, except for the large effective radius range (i.e., $\log R_{\rm H,eff} \gtrsim 0.6$) where the statistical trend for the weak and strong D4000 groups is not clearly seen mainly due to fewer number of sample spheroids. 

If the difference in the residuals to the fitting line between the weak and strong D4000 groups primarily originated from the difference in their light profiles, a possible explanation for the trend shown in the weak D4000 spheroids lying below the fitting line and the strong D4000 spheroids lying above the fitting line in surface brightness is that the strong D4000 spheroids generally have more concentrated and dense stellar light profiles than the weak D4000 spheroids, resulting in smaller (brighter) surface brightness (luminosity) in the strong D4000 strength spheroids. This explanation is consistent with the results shown in Figures \ref{fig8} and \ref{fig9} where the same strong D4000 spheroids tend to have larger S\'ersic index and $\Sigma_{\rm e}$ values.

Overall, the analysis performed in this section regarding the Kormendy relation for our sample of spheroids seems to support the idea that quenching of star formation activities is indeed related to the internal structure of galaxies \citep[e.g.,][at least apparently in spheroids]{bell12,lang14}.

\section{Discussion}
\label{sec:discussion}

\subsection{Morphological dependence on the ``downsizing'' trend}
\label{subsec:morphological dependnce on the downsizing trend}
Our main result is that there is a morphological dependence of the ``downsizing'' trend, such that galaxies with a prominent bulge component show a stronger mass quenching trend than galaxies without a prominent bulge component, even at a fixed stellar mass (i.e., Figures \ref{fig4} and \ref{fig6} and Table \ref{tab2}).  This suggests that the `mass' quenching is accelerated only after galaxies acquire a prominent bulge component. The acceleration of the mass quenching trend after the acquisition of a prominent bulge component in galaxies is likely to be due to either the role of a bulge component in preventing star formation activities, or the fading of a star-forming disk component, or both of the two effects working simultaneously.

\begin{table*}[ht]
\centering
\begin{threeparttable}
\caption{Mean p$-$values from the K$-$S tests for the pairs of the D4000 strength distributions between two selected morphological types}
\begin{tabular}{lllll}
\hline \hline
Morphology & Spheroids & Early-type Disks & Late-type Disks & Irregulars\\
\hline
Spheroids & 1.0 & 0.23 $\pm \ 0.21$ & $3.77\times10^{-6}$ $\pm \ 3.00\times10^{-5}$ & $4.81\times10^{-7}$ $\pm \ 3.51\times10^{-6}$\\
Early-type disks & $-$\tnote{a} & 1.0 & $0.006$ $\pm \ 0.016$ & $0.0003$ $\pm \ 0.0014$\\
Late-type disks & $-$ & $-$ & 1.0 & $0.04$ $\pm \ 0.08$\\
Irregulars & $-$ & $-$ & $-$ & 1.0\\
\hline \hline \\
\label{tab3}
\end{tabular}
{\small
\begin{tablenotes}
\item[a] The `$-$' mark indicates that the corresponding value is already given in the symmetric column and row.
\end{tablenotes}
}
\end{threeparttable}
\end{table*}

The role of a bulge component in preventing subsequent star formation activities in galaxies has been extensively discussed in previous studies \citep[e.g.,][]{deke09,sain12,bluc14,lang14,tacc15,huer16} where the gravitational potential of a prominent bulge component can stabilize the gas disk, preventing the gas disk from forming stars. At least, the quenching mechanisms associated with the prominence of a bulge component seem to exist in our sample of galaxies as well, as most of our quenched galaxies are bulge-dominated systems as shown in Figure \ref{fig3}.

Additionally, the presence of star-forming blue spheroids among our sample (i.e., spheroids that are outside the quiescence region in the $UVJ$ diagram of Figure \ref{fig3} and have the D4000 strength less than 1.5) supports the idea that the emergence of a prominent bulge component precedes the departure from the star-forming populations \citep[e.g.,][]{lang14}. The presence of blue spheroids at similar redshifts ($0.7 < z < 1.3$) has been also reported by \cite{powe17}, where the most likely mechanism responsible for the formation of blue spheroids was mentioned to be major mergers of star-forming disk galaxies. If most blue spheroids indeed originated from major mergers of star-forming disk galaxies, it would be interesting to note that most blue spheroids are relatively low mass systems (i.e., log$(M_{*}/M_{\odot}) \lesssim 10.0$).  This is seen in Figure \ref{fig4} and Table \ref{tab2} for our sample of blue spheroids, and from Figure 7 for those in \cite{powe17}. For these less massive systems, mergers that they might encounter would be likely to be major mergers for them (that is, merger mass ratio less than 4:1) due to their relatively low masses, which could possibly transform their morphologies into spheroids. Once they obtain a spheroidal morphology from the major mergers, the dominance of a bulge component-related quenching mechanisms such as gravitational stabilization of gas disks to prevent forming stars \citep[e.g,.][]{mart09,sain12} and the energetic active galactic nuclei (AGN) feedback triggered by major mergers \citep[e.g.,][]{spri05,hopk07} would be likely to occur, turning blue spheroids into red and quiescent spheroids and accelerating their mass quenching trend likely along with the associated size-mass relation for quiescent galaxies.

However, we also note that there are additional mechanisms that could possibly contribute to the formation of blue spheroids other than major mergers of star-forming disk galaxies. For instance, the `rejuvenation' of red spheroids by an external source could make the red spheroids blue again \citep[e.g.,][]{sali10}. In addition, environmental effect is one of the candidates that could contribute to the formation of blue spheroids \citep{lope16}. Although a single mechanism can not be pinned down for explaining the formation of blue spheroids \citep{toje13}, the tendency towards a high asymmetry index in blue spheroids reported by several studies \citep[e.g.,][]{lope16,powe17} seems to favor that most of blue spheroids come from merger-related mechanisms.

Late-type disks in our study, however, do not seem to show such a strong mass quenching trend as spheroids do (that is, Figure \ref{fig4} and Table \ref{tab2}). The main reason for late-type disks not to show the strong mass quenching trend is likely the dominance of mostly star-forming disk components in them, at least at the redshift of interest (i.e., 0.6 $\lesssim$ \lowercase{$z$} $\lesssim$ 1.2). The blue (thus, star-forming) disk components might resist against the mass quenching trend while the parent galaxies continue to grow in mass, unless they experience significant changes in their evolutionary paths such as additional gas supply or major mergers that could disrupt their disk components and consequently turn them into bulge-dominated systems morphologically. The resistance of star-forming disk components against the mass quenching mechanism seems to be related to what is called `slow quenching' where gas exhaustion in late-type galaxies occurs slowly over several Gyr. \citep[e.g.,][]{scha14}. Assuming that the `slow quenching' mechanism also works for our sample of late-type disks, they are expected to evolve into either massive red late-type disks or early-type disks (in case they have grown their bulge components more efficiently) unless they have experienced abrupt changes.

Although further analysis such as environmental effects, AGN activity, and structural parameters from bulge$-$disk decomposition are required to draw firm conclusions on the evolution of late-type disks, what is seen in our results (e.g., Figure \ref{fig4}, Table \ref{tab2}, and Section \ref{subsec:D4000 vs. stellar mass and surface density}) for them is that they do not tend to follow a strong mass quenching trend as bulge-dominated systems do, most likely due to the dominance of their star-forming disk components. 

Lastly, we note that {\it some} quiescent late-type disks do exist in our sample, although they are a significantly smaller fraction of the population than their star-forming counterparts, as previously reported \citep[e.g.,][]{mcgr08,bell12}.    Using D4000 $\geq 1.5$ as a criterion, there are 22 quiescent galaxies among 133 late-type disk galaxies (i.e., 16.5 \%), as given in Table \ref{tab2}. Dust reddening can increase the D4000 measurements for late-type disks that are near edge-on viewing angles. If we use an axis ratio $<0.5$ as a marker of edge-on geometry, and eliminate such inclined galaxies from our quiescent late-type disk sample, there are still 7 quiescent late-type disks (i.e., 5.3 \% of the full late-type disk sample; see Section \ref{subsec:D4000 vs. stellar mass and surface density} for details).  A similar fraction of late-type disks (4 of 133 galaxies, or 3.0 \%) are identified as quiescent using the criterion sSFR $< 10^{-10} \rm year^{-1}$. Regarding the possible mechanisms responsible for quiescent late-type disks, external processes such as environmental effects are often considered \citep[e.g.,][]{bass13}. Further analysis on environmental properties of quiescent late-type disks at intermediate redshifts is expected to shed some light on the formation mechanisms of these rare population of quiescent late-type disks.

While spheroids and late-type disks show distinct trends (including the mass quenching trend) in the correlations between stellar properties and galaxy structural properties explored in this study, early-type disks show intermediate trends that are between spheroids and late-type disks. This sounds obvious as the morphological selection of early-type disks is designed to do so, considering both the fractions $f_{\rm sph}$ and $f_{\rm disk}$ to be larger than 2/3 (see Section \ref{subsec:morphology} for details). Regarding this, it is interesting to note that even among early-type disks there are visually-identifiable morphological differences between moderately massive star-forming (that is, log$(M_{*}/M_{\odot}) > 10.0$ and sSFR $\gtrsim 10^{-10} \rm year^{-1}$) and massive quiescent galaxies (that is, log$(M_{*}/M_{\odot}) > 10.0$ and sSFR $< 10^{-10} \rm year^{-1}$), as previously mentioned in Section \ref{subsec:sSFR vs. stellar mass and stellar surface density}. In other words, massive star-forming early-type disks tend to have blue star-forming disk components, while the massive quiescent counterparts show more spheroid-like morphologies. This trend seems to suggest that morphological transformations in early-type disks gradually occur from disk-dominant systems to bulge-dominated systems as they become older and more massive.

Irregulars in our study overall seem to show the well-known properties such as relatively young stellar populations that are revealed by their colors, D4000 strength, and sSFR distributions and low concentration of light profiles as shown in their S\'ersic index and $\Sigma_{\rm e}$ distributions as previously reported \citep[e.g.,][]{oh13,eale17}. The mass quenching trend in irregulars does not seem significant, yet is slightly stronger in moderately high stellar mass regime (that is, log$(M_{*}/M_{\odot}) \gtrsim 10.0$) than that in late-type disks based on the correlation between the D4000 strength and stellar mass in Figure \ref{fig4} and Table \ref{tab2}. The slightly stronger mass quenching trend in irregulars in the moderately high stellar mass regime, however, is likely to be attributed to the presence of two different types of irregulars (i.e., typically young and star-forming clumpy irregulars and galaxy interaction-driven irregulars) in our total sample of irregulars as discussed in Section \ref{subsec:D4000 vs. stellar mass and surface density}, rather than their intrinsic ``downsizing'' trend.   

Regarding the evolution of irregulars, the well-known sequence of average stellar properties (e.g., age and metallicity) of galaxies along the Hubble sequence detailed in the local Universe \citep[e.g.,][]{oh13,khim15} and the decrease in the fraction of irregulars at least since $z \sim 2$ (\cite{huer16}) seem to suggest that the majority of irregulars have undergone morphological transformations from clumpy irregulars to structured late-type disks over time as long as they remain relatively undisturbed. Numerical simulations of the early evolution of disk galaxies performed by \cite{nogu99}, for instance, show that clumpy structures in a galactic disk can evolve to organized bulge and disk components likely into be late-type disks.

Our discussion in this section on the morphological dependence on the ``downsizing'' trend is further supplemented by the $p$-values from the K$-$S test for the D4000 strength distributions between spheroids, early-type disks, late-type disks, and irregulars. The associated $p$-values are tabulated in Table \ref{tab3}. In order to confirm that the $p$-values derived from the pairs of the D4000 strength distributions among morphologies statistically exist, we performed 1,000 K$-$S tests by considering the D4000 measurement errors to be the Gaussian random errors each time. Therefore, the $p$-values shown in Table \ref{tab3} are the mean $p$-values of the 1,000 different K$-$S tests, and the associated standard deviations are also given in the table. As shown in the table, it is interesting to note that all of the $p$-values are sufficiently low (that is, compared to the typically-employed $p$-value of 0.05) to suggest that the D4000 strength distributions (and thus, likely the overall age distributions of constituent stellar populations of galaxies) of the four morphological types are statistically different from one another. The only exception is the case between spheroids and early-type disks with the associated $p$-value of 0.23. The case between spheroids and early-type disks is likely to be related to their morphological similarity of being bulge-dominated systems.

Although the results from the K-S test seem interesting by themselves, they should not be regarded as evidence for a morphological dependence on the ``downsizing'' trend because the K-S test does not consider the stellar mass distribution of each morphological type. Rather, the results from the C-vM test for the correlation between the D4000 strength and stellar mass shown in Section \ref{subsec:D4000 vs. stellar mass and surface density} could indicate the morphological dependence on the ``downsizing'' trend since the C-vM test considers both the D4000 strength and the stellar mass distributions simultaneously. Recalling the results from the C-vM test shown in Section \ref{subsec:D4000 vs. stellar mass and surface density}, both results from the C-vM test and the K-S test show qualitatively similar trends, i.e., all pairs of the two selected morphological types show statistically different distributions except for the case between spheroids and early-type disks with the associated \textit{p}-values of 0.16 and 0.23 for the C-vM test and the K-S test, respectively. If the morphological selection is the only criterion applied to separate galaxies into the four morphological types, the results from the C-vM test shown in Section \ref{subsec:D4000 vs. stellar mass and surface density} for the correlation between the D4000 strength and stellar mass would be a piece of supporting evidence for a morphological dependence on the ``downsizing'' trend indeed.

\subsection{Morphological dependence on concentration of light profiles}
\label{subsec:Morphological dependence on concentration of light profiles}
As discussed in Sections through \ref{subsec:D4000 vs. stellar mass and surface density}, \ref{subsec:sSFR vs. stellar mass and stellar surface density}, and \ref{subsec:the concentration of galaxy light profiles}, the concentration of galaxies' light profiles (as explored by the S\'ersic index and $\Sigma_{\rm e}$ in this study, and often by the stellar surface density within the central 1 kpc in other works) is related to the stellar properties of galaxies. Quiescent galaxies tend to have more concentrated light profiles, indicating the predicting power of identifying quiescent galaxies through concentration \citep{fran08,fang13,oman14,willi17}.

Regarding the correlation between the concentration (equivalently, compactness) of light profiles of galaxies and quiescence, in addition to the possible physical mechanisms to explain the correlation, we also note that there is another possible explanation named the ``progenitor effect'' \citep{lill16,abra18}, which is somewhat of the ``progenitor bias'' \citep{vand96} in terms of selecting sample of galaxies over a span of cosmic time. This ``progenitor effect'' shows that the correlation between the central surface density and galaxy sSFR is naturally reproduced by assuming the evolution of the observed size-mass relation for star-forming galaxies and the appropriate mass-quenching trend \citep{peng10,caro13}, rather than invoking any physical link between the two parameters (that is, Figure 5 in \cite{lill16}).

Although both the possible physical mechanisms and the ``progenitor effect'' can explain the strong correlation between the concentration of light profiles of galaxies and the quiescence, it is clear that most of quiescent galaxies (i.e., D4000 $> 1.5$) with high concentration (i.e., S\'ersic index $ \gtrsim 4 $ and $\Sigma_{\rm e} \gtrsim 10^{9}$ $M_{\odot}\rm{kpc^{-2}}$) are morphologically bulge-dominated systems in our study (Figures~\ref{fig5}, \ref{fig7}, \ref{fig8}, and \ref{fig9}). Thus, it seems reasonable to assume that mechanisms responsible for higher concentration of light profiles are involved with visually-identifiable morphological transformations into bulge-dominated systems, likely either by the bulge growth or the fading of a disk component, or both effects working simultaneously.

In this regard, other than the discussion on the bulge growth in Section \ref{subsec:morphological dependnce on the downsizing trend}, several studies mainly with low$-z$ ($0.03 < z < 0.11$) galaxy samples \citep[e.g.,][]{vulc15,caro16} have suggested that the fading (or removal) of a disk component is the main cause for galaxies to be morphologically transformed into bulge-dominated systems, rather than the bulge growth itself. In particular, \cite{caro16} quantitatively demonstrated how the (either uniform or differential) fading of a disk component could make a bulge to total light ratio (B/T) increase without invoking actual bulge growth.  (for instance, B/T can increase from 0.4 to 0.63 with the fading of a disk component by $\sim$ 1 mag for an ordinary star-forming galaxy.)

Although our morphological analysis in this study alone cannot pin down which mechanism (that is, the bulge growth or the fading of a disk component) is more important than the other for morphological transformations into bulge-dominated systems, what is shown here is that the correlation between the concentration of light profiles of galaxies and the quiescence is involved with visually-identifiable morphological transformations of galaxies such that only bulge-dominated systems tend to have higher light concentration profiles with the signature of quiescence, since late-type disks and irregulars do not show such higher concentration of light profiles with the quiescence.

\section{Summary}
\label{sec:summary}
In this study, we have explored galaxy morphology and the associated stellar properties of galaxies in the redshift range $0.6 \lesssim z \lesssim 1.2$ by utilizing visual morphology, photometric, and spectroscopic information taken together. We have identified both spectroscopically and photometrically `red and dead' galaxies in the $UVJ$ diagram and further found that most ($\sim$ 97 \%) of these quiescent galaxies are morphologically bulge-dominated systems. 

Also, we have investigated a morphological dependence in the mass quenching trend in the diagram of the D4000 strength vs. stellar mass. In this diagram, we find a morphological dependence such that bulge-dominated systems show more significant mass quenching compared to disk-dominated systems within the same stellar mass range (i.e., Figure \ref{fig4} and Table \ref{tab2}). 
This trend is most likely due to the dominance of a blue star-forming disk component in late-type disks that would dilute the mass quenching trend compared to the bulge-dominated counterparts.

The correlations between the D4000 strength and galaxy structural parameters such as the S\'ersic index and the stellar surface density within the effective radius (which is often considered to be related to the quiescence), $\Sigma_{\rm e}$, show that only some fraction of the bulge-dominated galaxies have large D4000 strength (D4000 $>$ 1.5), $\Sigma_{\rm e}$ ($\Sigma_{\rm e} \gtrsim 10^{9}$ $M_{\odot}\rm{kpc^{-2}}$), and S\'ersic index (S\'ersic index $\gtrsim$ 4), indicating that not all of the bulge-dominated galaxies have high concentration light profiles with the signature of quiescence. On the other hand, the late-type disks and irregulars generally show neither such highly concentrated light profiles nor signatures of quiescence.

A morphological analysis on the correlations between sSFR, stellar mass, $\Sigma_{\rm e}$, and the S\'ersic index basically shows qualitatively similar trends to those based on the D4000 strength. In particular, in the diagram of sSFR vs. stellar mass, bulge-dominated systems such as spheroids and early-type disks are clearly separated into the star-forming sequence and the quiescent sequence, while late-type disks and irregulars within the same stellar mass range considered have formed the star-forming sequence only, which is likely due to the star-forming disk or clumpy components in them. 

The difference in the concentration of light profiles between weak and strong D4000 galaxies in each morphological type is shown to be the most significant in spheroids and the least significant in late-type disks (i.e., Figures \ref{fig8} and \ref{fig9}). The Kormendy relation for spheroids further shows that at fixed effective radius strong D4000 spheroids tend to have smaller (brighter) surface brightness (luminosity) than weak D4000 counterparts by $\sim$ 0.8 mag on average, suggesting more concentrated light profiles in strong D4000 strength spheroids at fixed effective radius. Although the correlation between the high concentration light profiles and the quiescence in spheroids could be alternatively explained by the combination of the observed evolution of the size-mass relation for star-forming galaxies and the ``progenitor'' effect of quiescent populations \citep[i.e.,][]{lill16}, the morphological analysis regarding this correlation in our study suggests that most quiescent and centrally dense galaxies at intermediate redshifts at least have a visually-identifiable prominent bulge component and that none of late-type disks and irregulars show the high concentration light profiles with the signature of quiescence.

Taken together, our analysis of galaxy structure (including visual morphology) and stellar properties of galaxies through photometric and spectroscopic information as well as galaxy structural parameters indicates that physically-motivated quenching mechanisms are required to appropriately explain the morphological behavior shown in the various diagrams explored in this study. Observationally, future studies with significantly larger sample of galaxies are expected to enable us to better understand the morphology-entangled galaxy properties and the associated quenching mechanisms at intermediate redshifts statistically. All in all, our findings suggest a morphological dependence of the downsizing trend, and a tight correlation between quenching and the presence of a prominent bulge component, such that most of quenched galaxies at intermediate redshifts are bulge-dominated systems.

\acknowledgments
We thank the scientific editor and the referee for their constructive comments that helped improve the paper significantly. We thank the US National Science Foundation for its financial support through grant AST-1518057. This work has been supported by HST-GO-13779 from STScI, which is operated by the Association of Universities for Research in Astronomy, Inc., for NASA under contract NAS 5-26555.
\clearpage

\clearpage
\end{document}